\documentclass[useAMS,usenatbib]{mn2e}

\usepackage{graphicx}
\usepackage{natbib}
\usepackage{subfloat}
\usepackage{amssymb}
\usepackage{amsbsy}
\usepackage{amsmath}
\usepackage{color}
\usepackage[normalem]{ulem} 

\def\dch#1{{#1}}

\def\alv#1{{#1}}

\def\kch#1{{#1}}
\def\kco#1{{#1}}

\def\dchg#1{{#1}}

\title[An Approximate Analytic Model of a Star Cluster with Potential Escapers]{An Approximate Analytic Model of a Star Cluster with Potential Escapers}
 \author[Kathryne J. Daniel, Douglas C. Heggie, and Anna Lisa Varri]{Kathryne J. Daniel$^{1,4}$\thanks{E-mail: {kjdaniel@brynmawr.edu} (KJD); d.c.heggie@ed.ac.uk (DCH); varri@roe.ac.uk (ALV)}, Douglas C. Heggie$^{2}$, 
and Anna Lisa Varri$^{3}$
\\ 
$^{1}$Department of Physics, Bryn Mawr College, 101 N. Merion Avenue, Bryn Mawr, PA 19010, USA \\
$^{2}$School of Mathematics and Maxwell Institute for Mathematical Sciences, University of Edinburgh, King's Buildings, Edinburgh EH9 3FD, UK\\
$^{3}$Institute for Astronomy, University of Edinburgh, Royal Observatory, Blackford Hill, Edinburgh EH9 3JZ, UK\\
$^{4}$Department of Physics \& Astronomy, Johns Hopkins University, 3400 North Charles Street, Baltimore, MD 21218, USA}

\def\gtorder{\mathrel{\raise.3ex\hbox{$>$}\mkern-14mu
             \lower0.6ex\hbox{$\sim$}}}
\def\ltorder{\mathrel{\raise.3ex\hbox{$<$}\mkern-14mu
             \lower0.6ex\hbox{$\sim$}}}
\def\be{{\mathbf e}}
\def\bexr{{\mathbf e}_{xR}}
\def\bexn{{\mathbf e}_{xN}}
\def\beyr{{\mathbf e}_{yR}}
\def\bjn{\mathbf{J}_N}
\def\bomega{{\boldsymbol{\omega}}}
\def\br{{\mathbf r}}
\def\bronen{{\mathbf r}_{1N}}
\def\brn{{\mathbf r}_N}
\def\bron{{\mathbf r}_{0N}}

\def\brr{{\mathbf r}_R}

\def\bv{{\mathbf v}}
\def\bvn{{\mathbf v}_N}
\def\bvr{{\mathbf v}_R}
\def\hbar{{\langle H_K\rangle}}
\def\hcrit{H_{crit}}
\def\hopt{H_{opt}}
\def\jn{J_N}
\def\jnz{J_{zN}}
\def\jzn{\jnz}
\def\jzopt{J_{z,opt}}
\def\nbody{$N$-body}
\def\Omegao{\Omega_0}

\def\rmax{r_{max}}
\def\rt{r_t}
\def\xor{x_{0R}}
\def\yor{y_{0R}}
\def\xn{x_N}
\def\xr{x_R}
\def\yr{y_R}
\def\S{Section }
 
\date{Accepted 2017 March 2. Received 2017 March 1; in original form 2016 September 5}

\pagerange{\pageref{firstpage}--\pageref{lastpage}} \pubyear{2017}

\begin{document}

\maketitle

\label{firstpage}


\begin{abstract}
In the context of a star cluster moving on a circular galactic orbit, a \lq\lq potential escaper" is a cluster star that has orbital energy greater than the escape energy, and yet is confined within the Jacobi radius of the stellar system. On the other hand analytic models of stellar clusters typically have a truncation energy equal to the cluster escape energy, and therefore explicitly exclude these energetically unbound stars. Starting from the landmark analysis performed by H{\'e}non of periodic orbits of the circular Hill equations, we present a numerical exploration of the population of ``non-escapers'', defined here as those stars which remain within two Jacobi radii for several galactic periods, with energy above the escape energy.   We show that they can be characterised by the Jacobi integral and two further approximate integrals, which are based on perturbation theory and ideas drawn from Lidov-Kozai theory.  Finally we use these results to construct an approximate analytic model that includes a phase space description of a population resembling that of potential escapers, in addition to the usual bound population.
\end{abstract}




\section{Introduction}\label{Sec:Introduction}

\subsection{Tidal effects on the structure of star clusters}

As a star cluster orbits around the centre of the host galaxy, the tidal forces induced by the galactic potential {shape the} 
``unstable'' orbits on which stars, in principle, can escape from the cluster. In the simplest case, we may consider the star cluster as moving on a circular orbit around the galactic centre, and the motion of a star in the potential of the galaxy-cluster system may be treated as a restricted three-body problem (for a review of ``Hill's problem'' in the context of star cluster dynamics, see \citealp{Heggie2001}).  Several of the interesting physical mechanisms that underlie the dynamical evolution of star clusters depend on the effects of the tidal field of the host galaxy. None the less, star clusters are still often studied in the context of simple analytic models in which the action of tides is implemented by imposing a suitable energy truncation, perhaps supplemented by a prescription for the escape of stars.

In this context a classical problem in stellar dynamics is the search for self-consistent solutions of the Boltzmann equation with a tidal cutoff, such as the family of models proposed by King (1966). These solutions are often defined as functions of constants of the motion which characterise a stellar system, e.g. the stellar energy. In this approach, the starting point is the identification of an appropriate form for the distribution function in phase space.  Indeed, as a zeroth-order dynamical description, the simple classes of spherical equilibrium models defined by a Maxwellian distribution function, suitably modified near the tidal boundary and truncated above it (e.g. \citealp{WD61,King66,Wilson75} and, more recently, \citealp{Gomez14} and \citealp{GielesZocchi2015}), have had remarkable success in  reproducing the observed properties of globular clusters \citep[e.g. see][]{MvdM05}.

Of course, the choice of the tidal modification in the definition of the distribution function strongly affects the structural and kinematic properties of the resulting configurations, at least in the outer parts of the cluster \citep[see][]{Davoust77, Hunter77}. Modifications which give smoother distribution functions, such as in \cite{Wilson75} models, generally produce configurations with more extended halos, and these equilibria are often more successful than \cite{King66} models in reproducing the surface brightness and velocity dispersion profiles of star clusters in the proximity of the prescribed truncation radius (see \S 4 of \citealp{MvdM05}, and other references listed below). This aspect is particularly relevant for the dynamical study of star clusters characterised by significant ``extra-tidal'' structures, i.e., with a surface brightness profile extending beyond the cut-off predicted by spherical King models (e.g. see the structural parameters of star clusters in the Milky Way and its satellites as determined by \citealp{MvdM05} and {\citealp{Miocchi2013}}, in M31 by \citealp{Barmby2002}, and in NGC 5128 by \citealp{Harris2002}, especially their table~3), or velocity dispersion profiles which are peculiarly flattened in the outskirts of the cluster (e.g. see the studies of M15 and M92 by \citealp{Drukier1998,Drukier2007}, $\omega$ Cen by {\citealp{Sollima09}} and \citealp{DaCosta2012}{, and NGC 5694 by \citealp{Bellazzini2015}}). {In this context, an heuristic approach to the description of the density profile in the proximity of the tidal boundary has often been adopted, by characterising its slope in terms of a power-law function (e.g. see \citealp{Grillmair1995, Leon2000}, and more recently \citealp{Olszewski2009, Correnti2011, Kuzma2016}) or similar templates (see \citealp{Elson1987,Kupper10}).} 

Such analytic equilibrium models {and templates} are usually limited to a very idealised treatment of the cluster-galaxy system, in which the tidal field is assumed to vanish within the Jacobi radius.  In contrast, its effects are included fully in the slightly more realistic equilibrium models of \citet[][]{HR1995} and \citet{BertinVarri2008}, but these models have not been applied to observed clusters, and cannot account for the observed flattened velocity dispersion profiles just inside the Jacobi radius. In fact, so far only numerical simulations by means of $N$-body codes, in which an external tidal field can be taken into account explicitly, provide a tool for the full study of the evolution of a tidally perturbed cluster, especially when elliptic galactic orbits are considered, so that tidal effects are time-dependent \citep[see][]{Aarseth03,RenaudGieles2015}. In particular, this approach has led to detailed investigations of the rich morphology and kinematics of the tidal tails, i.e. the streams of stars which have escaped from the cluster, resulting in a major improvement of our current understanding of these striking morphological and dynamical features \citep[e.g. see][]{Johnston99,Kupper08}.

In addition to the existence of tidal tails, numerical simulations have also shown that even star clusters on simple circular orbits possess a population of stars with energies above the escape energy which are {none the less} confined within the stellar system itself (``potential escapers'').  Though the definition sounds contradictory, it was realised long ago by \citet{Henon70} that such stars could exist.  They were later studied with $N$-body simulations by \citet{2001ASPC..228...29H} and \citet{2001MNRAS.325.1323B}. {From the theoretical point of view, potential escapers play a fundamental role in determining the properties of the process of escape from a star cluster (especially its time scale, see \citealp{Ross97}, \citealp{2000MNRAS.318..753F},  and related prescriptions by \citealp{Takahashi00}, \citealp{Giersz13}, and \citealp{Sollima14} for their implementation in the context of Fokker-Plank and Monte Carlo codes, respectively).} { From a more phenomenological perspective,} \citealp{Kupper10} have shown that such a population of energetically unbound stars dominates the mass distribution of a star cluster above about 50 per cent of its Jacobi radius, and that beyond 70 per cent nearly all stars are potential escapers. This investigation also revealed that the behaviour of the main observables, especially the surface brightness and velocity dispersion profiles, in the proximity of the tidal limitation of the system, is
almost entirely shaped by the contribution of these energetically unbound stars.          

This recent progress on the numerical exploration of the structural and kinematic properties of tidally perturbed star cluster models is particularly relevant in the context of the forthcoming ``era of precision astrometry'' for Galactic studies. The exquisite astrometric information  {which is beginning to emerge from} 
the Gaia mission, combined with ground­-based wide­-field imaging, archival information from the Hubble Space Telescope, and detailed spectroscopic campaigns (e.g. Gaia­-ESO survey, see \citealp{Gilmore2012}), will allow us to access, for the first time, virtually the full phase space of several Galactic star clusters. In this respect, a key element is that such a richness of observational data will allow us to map stars in the outer regions of several Galactic globular clusters on the basis of  photometry, proper motions and parallaxes, leading to a proper separation of cluster members and foreground/background stars; such a distinction is critical for studying the outskirts of globular clusters, especially to empirically distinguish between energetically bound and unbound stars. 

Unfortunately, this growing body of numerical and observational information is not matched by comparable progress in the theoretical understanding of the phase space properties of tidally perturbed stellar systems, and, as a result, none of the analytic models which are currently available include the contribution of the potential escapers. The need of such a tool is the stimulus that triggered the present study. 

\subsection{Outline of the paper}

{We begin by reviewing Hill's equations (Sec.\ref{sec:units}) and their invariant $\Gamma$ (the Jacobi integral).  {These are a special form of the equations of motion of a star in a cluster on a circular galactic orbit, and} 
form the basis of H{\'e}non's 
landmark analysis 
of {a remarkable family} 
of {stable, planar, periodic} orbits 
\citep{Henon69,Henon70}, which we review in {Sec.\ref{sec:fandg}}.  {Their particular relevance for this paper is that they contain easily understood examples of potential escapers.}  Sec.\ref{mapeq} extends H\'enon's survey into a numerical exploration  of three-dimensional orbits.  The initial conditions are still chosen by an artificial procedure, and so Sec.\ref{sec:invariantf}, which is really preparation for later sections of the paper, discusses how to select initial conditions according to principles of equilibrium statistical mechanics (the ``microcanonical ensemble'').}

{ Sec.\ref{Sec:} begins with theoretical and numerical results on the dynamics of orbits above the escape energy, and then (Sec.\ref{sec:empirical}) sets up a criterion for approximately determining, on the basis of initial conditions alone, whether a particle escapes or not.  In this part of the study we use sets of orbits for various discrete values of $\Gamma$, each set using essentially a microcanonical distribution of initial conditions.
}  
{  Then in Sec.\ref{sec:pe-observables} a new data set is considered, with random, uniformly distributed values of $\Gamma$, for each of which the initial conditions are again selected from the microcanonical distribution.  The spatial and kinematic distributions of this data set are described.   Then Sec.\ref{Sec:delta} builds a composite model from a Woolley model plus a distribution of non-escapers constructed by applying the escape criterion to a {\sl canonical} distribution (in analogy with the distribution function of the Woolley model).   We conclude the paper by presenting an extensive discussion of the limitations and strengths of the resulting equilibria (\S\ref{Sec:DC}).   }

\subsection{A note on nomenclature}\label{sec:nomenclature}

{ The notion of potential escapers, which we have already defined and referred to several times, is quite well established in the literature, but it is not identical to the idea of non-escapers, with which we will be concerned in practice in this paper.  In an $N$-body model, potential escapers include transients, such as ejecta from three-body interactions, which leave the cluster promptly.  It also includes stars whose energy comes to exceed the escape energy simply because the potential well of the cluster becomes shallower as a result of  escape.  It includes stars whose energy has been altered by two-body relaxation.  None of these processes are included in our work, which considers orbits in a simplified gravitational field, which is smooth and static in a frame which rotates with the cluster about the galactic centre.   In such a model there are stars which never escape, and others for which the time scale of escape can be very extended.  In the context of numerical integrations, these are indistinguishable, and so we are obliged to define ``escape'' in an essentially arbitrary but practical manner which relates to the astrophysical context.  }

{We now describe the choice we have made, leaving more detailed discussion to Sec.\ref{Sec:int-time}.  
  Throughout this investigation, we operationally define a ``non-escaper'' as an orbit, at an energy above the energy of escape, such that its maximum radius, during a time of 8 revolutions of the cluster around the galaxy
, is less than twice the Jacobi radius.  The criterion which we construct in Sec.\ref{sec:empirical} aims to match the distribution of non-escapers, defined in this sense, and so a star on an  orbit satisfying this criterion  
  will be referred to as a ``predicted non-escaper''.  }

\begin{figure}
\begin{center}
\centerline{\hspace{-1.3cm}\includegraphics[width=0.38\textwidth]{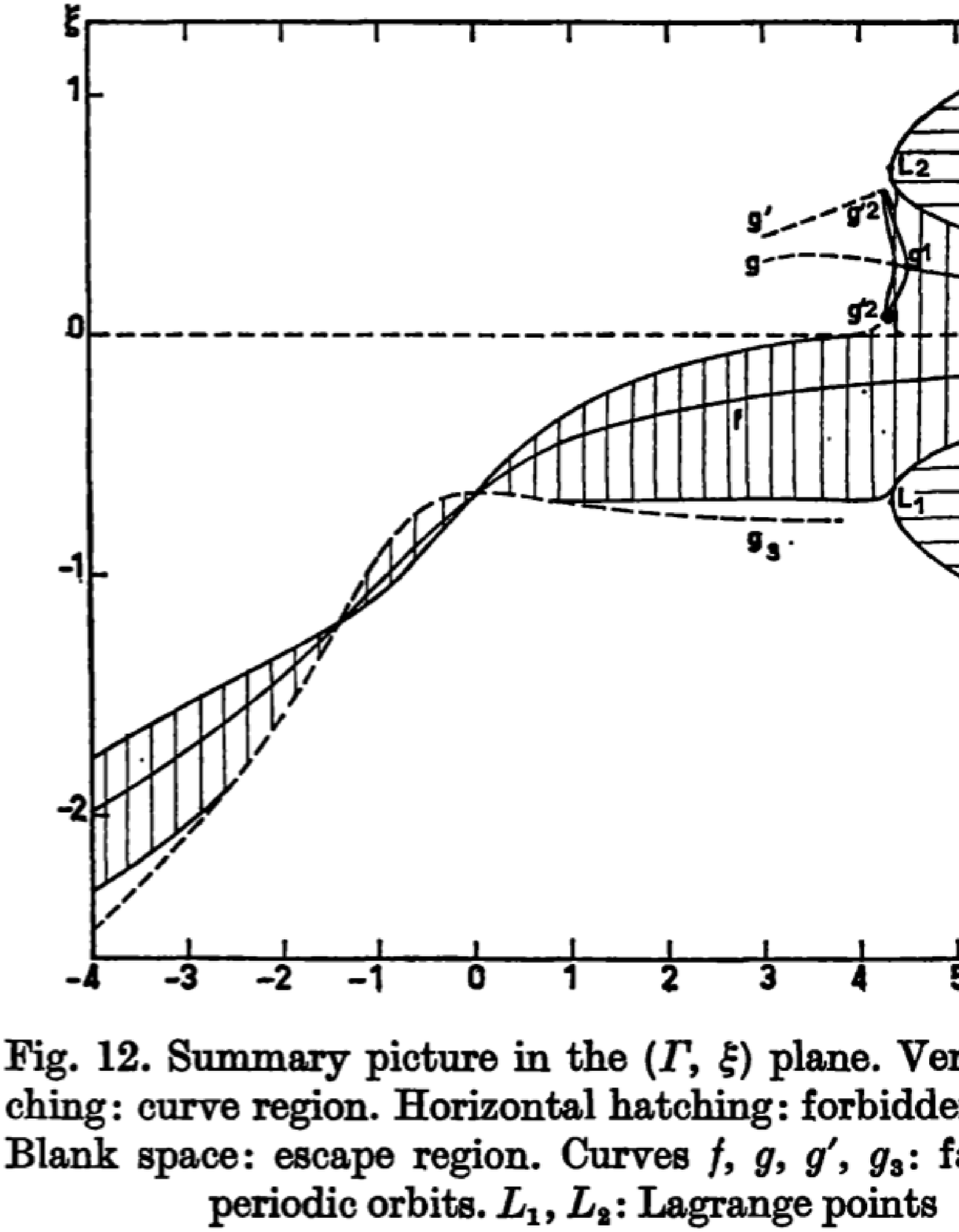}}
\caption{Figure~12 from \citet{Henon70}.  The horizontal axis shows the 2D analogue to the Jacobi integral, $\Gamma$, while the vertical axis shows the $x_R$-coordinate of the starting point, which H\'enon calls $\xi$. The curve  marked \lq\lq f" gives the initial conditions for the $f$-orbit family. Credit: M. H\'enon, A\&A, 9, 30, 1970, reproduced with permission \textcopyright ESO.}
\label{fig:henonfig12}
\end{center}
\end{figure}


\section{Numerical studies of non-escaping orbits in Hill's problem}
\label{Sec:Theory}

\subsection{Equations of motion}
\label{sec:units}

In a linear approximation of the tidal field, the equations of motion of a star in a star cluster are given in \citet{Chandra}. The coordinate system \dch{he adopts} has the origin placed at the centre of the globular cluster and the frame co-rotates with the cluster. The Cartesian unit vector $\bexr$ points radially away from the galactic centre, and $\beyr$ points in the direction of rotation about the galactic centre, where the subscript \lq\lq R" indicates this rotating frame. We use units in which $G=1$, the cluster mass is unity and the angular velocity of motion about the galaxy is also unity. Thus the unit of time at the sun's distance from the Galactic Centre would be about $3.5\times10^7$yr. If we assume furthermore that the galactic and cluster potentials are Keplerian, then the equations of motion are those of Hill's problem, i.e.
\begin{eqnarray}
\ddot{x}_R&=&2\dot{y}_R+3x_R-\dfrac{x_R}{r_R^3},\label{eqn:EOMx}\\
\ddot{y}_R&=&-2\dot{x}_R-\dfrac{y_R}{r_R^3}\label{eqn:EOMy}\\
\ddot{z}_R&=&-z_R-\dfrac{z_R}{r_R^3}\label{eqn:EOMz}.
\end{eqnarray}
where
\begin{equation}\label{eqn:r3D}
r_R = (x_R^2+y_R^2+z_R^2)^{1/2}.
\end{equation}

Throughout our analysis it will also be convenient to use a coordinate system which shares the same origin as the above rotating frame, i.e. at the centre of the cluster potential, but is non-rotating. The two coordinate systems coincide at $t=0$, and, since the angular velocity of the cluster around the galaxy is unity, the components of the velocity are given by
\begin{eqnarray}
\dot{x}_N &=& \dot{x}_R \cos t - \dot{y}_R \sin  t - y_N \\
\dot{y}_N &=& \dot{x}_R \sin  t + \dot{y}_R \cos  t + x_N \\
\dot{z}_N &=& \dot{z}_R,
\end{eqnarray}
where the subscript \lq\lq N" indicates the non-rotating frame. The corresponding equations of motion are
  \begin{equation}
    \ddot{\br}_N = - \dfrac{\brn}{r^3} +2x_R\be_{xR} - y_R\be_{yR} - z_R\be_{zR},\label{eq:EOMN}
  \end{equation}
  where $\be_{xR},\be_{yR},\be_{zR}$ are the three unit vectors of the rotating frame. This way of expressing the tidal acceleration is convenient because the galactic tidal potential is constant in the rotating frame. (See equation~\ref{eq:phit} below.)

These equations have an integral, the Jacobi integral, which will be very important throughout this study, and is defined as 
\begin{equation}\label{eqn:Gamma3D}
\Gamma=3x_R^2+\dfrac{2}{r_R} -z_R^2 -\dot{x}_R^2 -\dot{y}_R^2 -\dot{z}_R^2 = -2E_R,
\end{equation}
\dch{where}
$E_R$ is the energy in the rotating frame. In terms of the non-rotating coordinate system, the energy and the Jacobi integral can be related in the form
\begin{equation}
H_K = -\frac{1}{2}\Gamma + J_{zN} - \Phi_t,
\label{eq:gamma}
\end{equation}
{ in which the new notation is defined as follows: }
  \begin{itemize}
  \item 
{ $H_K$ is  the Keplerian energy in the non-rotating frame, i.e.}
\begin{equation}
H_K = \displaystyle{\frac{1}{2}\bvn^2 - \frac{1}{r}},    
\label{eq:hk}
\end{equation}
{where $\bvn$ denotes the velocity vector};
\item  $J_{zN}$ is the $z$-component of the angular momentum in the non-rotating frame, which can be written as
\begin{equation}
J_{zN} = \bomega.(\br_N\times\bvn),\label{eq:jzn}
\end{equation}
where $\bomega$, the angular velocity of the cluster around the galaxy, is simply the unit vector in the $z$-direction; and
\item    $\Phi_t$ is the tidal potential, which is most economically expressed in the rotating frame, i.e.
\begin{equation}
\Phi_t = -x_R^2 + \frac{1}{2}y_R^2 + \frac{1}{2}z_R^2\label{eq:phit}.
\end{equation}
  \end{itemize}


\subsection{H\'enon's $f$ orbital family}
\label{sec:fandg}

During his exploration of the restricted 3-body problem in the 2D Hill's approximation, \cite{Henon69,Henon70} identified a family of stable periodic orbits, which he called family $f$ and which have formed the starting point of our own explorations. In those papers his exploration was limited to planar orbits which start on the $x_R$-axis with $\dot x_R = 0$; then $\dot y_R >0$ is determined by the value of $\Gamma$.

For the reader's convenience, Fig.~\ref{fig:henonfig12} shows a re-print of fig.~12 from \cite{Henon70}. The horizontal axis shows the Jacobi integral $\Gamma$, and the vertical axis shows the initial value of $x_R$. The horizontal hashed regions are \lq\lq forbidden" in the sense that their boundary marks the zero velocity curves of the effective potential.  A star with $\Gamma<\Gamma_J\equiv 3^{4/3}\simeq 4.33$, will have energy greater than the critical energy at the Lagrange points $L_1,L_2$, which are marked on the figure; they lie at $x_R = \pm r_J$, where the Lagrangian (Jacobi) radius is  $r_J \equiv 3^{-1/3}\simeq 0.69$.  Initial conditions for the $f$-orbital family are marked \lq\lq f".  Vertically hashed regions are regions of stability, in the sense that a star launched from position $x_R$ in the hashed region, with initial velocity defined as above, describes a closed invariant curve in a surface of section.  (See \citealt[][\S 3.2.2]{2008gady.book.....B} for an introduction to the use of this technique in stellar dynamics, though it plays no further role in our study.)  For our purposes the importance of these findings is that, in the planar problem considered by H\'enon, such a star must remain in the vicinity of the star cluster for all time.  {Indeed, in the sense used in this paper (see Secs.\ref{sec:nomenclature} and \ref{mapeq})} 
  it is a non-escaper, {provided that its radius remains less than $2r_J$ up to at least time $16\pi$}.

\begin{figure*}
\begin{center}
\includegraphics[width=\textwidth]{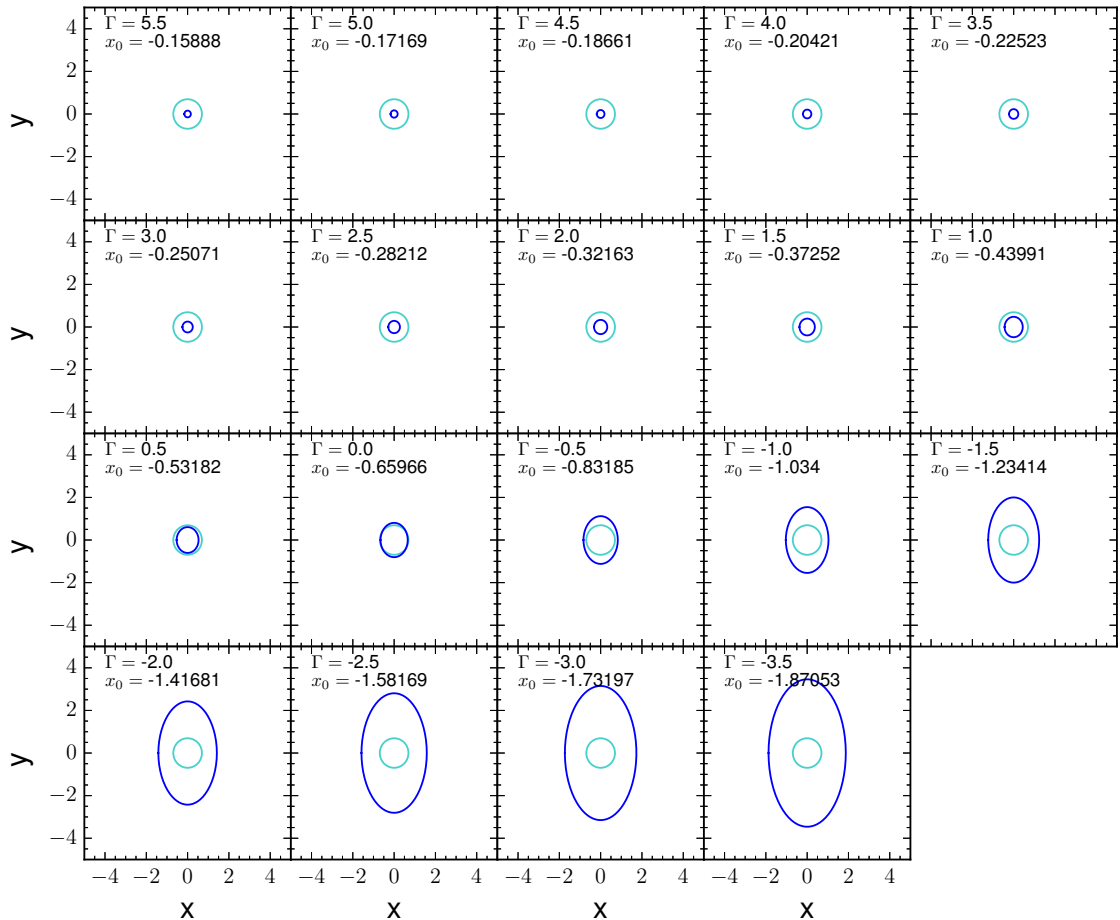}
\caption{Examples of stable $f$-orbits with initial conditions from \citet{Henon69}. The associated $\Gamma$ and initial position, $x_0\equiv x_R(0)$, for each orbit {are}  
  printed as an inset. The abscissa and ordinate are $x_R, y_R$, respectively.}
\label{fig:f}
\end{center}
\end{figure*}

\begin{figure*}
\begin{center}
\centerline{\includegraphics[width=1.1\textwidth]{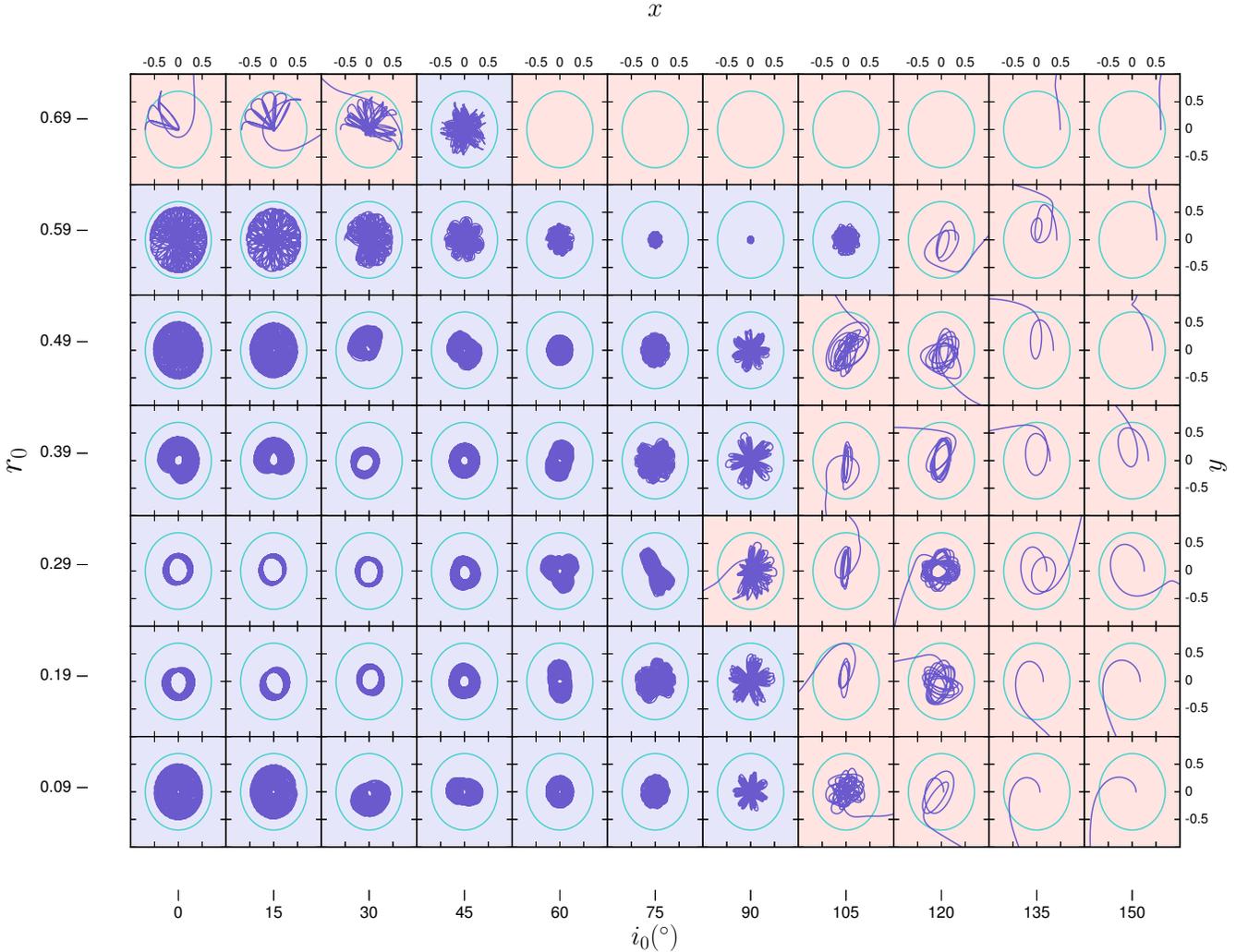}}
\caption{Three-dimensional orbits with $\Gamma=3$, projected onto the $x_N,y_N$ plane in the non-rotating frame.  At upper left is a planar orbit starting on the $\xr$-axis at $\xr = -r_0 = -r_J$ with a velocity orthogonal (in the rotating frame) to the $\xr$-axis. In general the initial position is $x_{0R} = - r_0\cos i_0, y_{0R} = 0, z_{0R} = r_0\sin i_0$. $\Gamma$ is held constant, while $r_0$ is decreased by $0.1$ with each step down, and $i_0$ is increased by $15^\circ$ with each step to the right. {This choice of initial conditions, for some values of the inclination, may be impossible (i.e., the initial velocity would be 
    imaginary,  
    see equation~9); in such a case, we have left the corresponding panel empty (see five occurrences in the top row).} The initial parameters $(r_0,i_0)$ for each orbit are given by the lower-left axis labels. The orbital trajectory in the non-rotating frame is plotted in purple, with the axes labelled upper-right. The tidal radius is shown as a cyan circle, distorted slightly by the scaling of the graphics. The classification of each orbit (as escaper or non-escaper) is determined empirically, as described in the text, where a red background indicates an escape orbit and a lavender background indicates a non-escaper. Orbits with $i_0 = 165^\circ$ and $180^\circ$ are not shown; they are escapers, and qualitatively similar to those in the last column. The integration time is $32\pi$.}
\label{fig:g3plus}
\end{center}
\end{figure*}

In Fig.~\ref{fig:f}, we show several examples of $f$-orbits where we have used the initial conditions published in table~3 of \cite{Henon69}.\footnote{We have used a variable-order integrator from the scientific Python integration library (odeint), and tested our orbits to ensure that the change in $\Gamma$ remains less than $10^{-6}$.} The orbit is shown in dark blue while the tidal (Jacobi) radius ($r_J$) is cyan.  $f$-orbits with low values of $\Gamma$ appear as approximate epicycles to orbits about the galactic centre with guiding centre at $x=y=0$.  At high values of $\Gamma$, $f$-orbits appear as approximate circular Keplerian orbits about the cluster. The orbits in this family cross from lying entirely within to entirely outside the Jacobi radius at a threshold value $\Gamma\approx 0$. For our purposes, the importance of these results is that they show us in simple terms that it is possible for a star to remain inside the cluster even though its energy exceeds the energy of escape (i.e. $\Gamma < \Gamma_J$).  It is also shown in Appendix \ref{sec:appb} that the $f$-orbits can be described successfully, in the range $\Gamma\gtorder0$, by using first-order perturbation theory; they are approximately Keplerian orbits with a tidal perturbation. This result was important in guiding us to the possibility that non-escaping orbits {\sl in general} might be thought of as tidally perturbed Keplerian motions.


\subsection{Two-parameter exploration}
\label{mapeq}   

We now expand H\'enon's exploration from a one{-parameter family of planar orbits}  
  (at fixed $\Gamma$) to {a two-parameter family of three-dimensional orbits}
, that is, we consider starting positions on the $x_R,z_R$ plane $y_R=0$.  As in \S\ref{sec:fandg}, the starting velocity is orthogonal to this plane, but henceforth we restrict values of $\Gamma$ to the range $[0,\Gamma_{J}]$, and in this subsection restrict the starting position to points within the Jacobi radius $r_J$.

In Fig.~\ref{fig:g3plus} we show a tabular sample of the orbits, projected onto the $x_N-y_N$ plane of the non-rotating frame. These show results for $\Gamma = 3$, and we have carried out similar visual studies for several other values of $\Gamma\in[0,\Gamma_J$]. The top, leftmost panel shows a planar orbit starting at $y_{0R} = 0, x_{0R} = - r_0 = - r_J$.  Thus it corresponds to the motion in Fig.\ref{fig:henonfig12} at $\Gamma = 3, \xi = -r_J$, close to the lower boundary of the vertically hatched region surrounding family $f$. Each row down has an initial radius, $r_0$, that incrementally decreases by $0.1$ (in the units described in \S\ref{sec:units}) and each column to the right has incrementally increasing initial inclination, $i_0$, from the $x_R-y_R$ plane (where $i_0=0$ is on the negative $x_R$-axis pointing toward the galactic centre at $t=0$ and increases via the positive $z_R$-axis towards the positive $x_R$-axis in steps of $15^\circ$).  The values for the initial radius and inclination are shown at left and bottom, respectively. We integrated each orbit for sixteen galactic orbital periods, i.e. $32\pi$ units, corresponding to approximately 3.5Gyr at the Sun's Galactocentric distance.

The colour of each panel divides the orbits into two classes, which we refer to as ``escaper'' and ``non-escaper''. Operationally, our definition is that an orbit is classed as a ``non-escaper'' if its maximum radius $r$, over an integration time of $16\pi$ (i.e. half the integration time in Fig.\ref{fig:g3plus}), is less than $2r_J$, and an escaper otherwise.  Our reason for choosing a radius larger than $r_J$ is the possibility, suggested by Fig.\ref{fig:henonfig12}, that there may exist permanently bound planar orbits with maximum distance slightly larger than $r_J$, even for values of $\Gamma$ in the restricted interval which we have explored.  This issue is discussed further in \S\ref{Sec:int-time}, though for the small sample of orbits in Fig.\ref{fig:g3plus} it may seem that this precaution is superfluous; non-escapers appear to remain within the Jacobi radius, though it must be recalled that we have here only the projection of three-dimensional motions. Similarly, there are no orbits classed here as ``non-escapers'' which  escaped over the longer integration time of $32\pi$.
Indeed, the majority of escapers were classified as such within the first few orbital periods. We further discuss our classification scheme in the context of integration time in \S\ref{Sec:int-time}.

A central aim of our efforts in \S\ref{Sec:} will be to build a simple criterion which enables us to distinguish escapers from non-escapers. However, in order to construct an equilibrium distribution function, such a criterion should depend only on integrals of the motion, which is the subject of \S\ref{sec:integrals}. All we see from the results shown in Fig.\ref{fig:g3plus} is that non-escapers in this sample generally start inside the Jacobi radius and are retrograde (i.e. $i_0 < 90$ in our notation).  The latter characteristic of non-escapers has been known for a long time (e.g. our Fig.\ref{fig:henonfig12}, and \citealt[][and references therein]{1975AJ.....80..290K}). These authors note that this finding may be understood in terms of resonance between the motion of the star and that of the cluster, but also observe that retrograde motion results in a Coriolis acceleration towards the cluster centre in the rotating frame, which thus effectively operates in such a way as to enhance gravitational attraction {(see also \citealt{Read06})}. In our explorations, at fixed $\Gamma$, an analogous effect also emerges from equation~(\ref{eq:gamma}).  This equation shows that retrograde motions in the non-rotating frame, which have $J_{zN}<0$, correspond to more negative values of the Keplerian energy $H_K$. Thus retrograde orbits have smaller semi-major axis (in the Keplerian approximation), which means that they are subject to smaller tidal acceleration, and are therefore less liable to escape.


\subsection{Full exploration at fixed $\Gamma$}
\label{sec:invariantf}

{
While orbits may be expected to cross the $x,z$ plane at some point, there is no reason why they should do so orthogonally. Therefore the results of the previous subsection by no means provide full coverage of phase space, even if this is restricted to a fixed value of $\Gamma$ and orbits inside the Jacobi radius. In Appendix \ref{sec:appa} we therefore present a different procedure to sample a $\Gamma$-hypersurface in phase space.}  It is based on the microcanonical ensemble of statistical mechanics, and therefore the distribution on a $\Gamma$-hypersurface would be time-invariant if the hypersurface were finite.  That is not the case here, however, and so we also impose a condition 
$r < r_{ic}$, {where the threshold value $r_{ic}$ is to be chosen.   We have explored values from $r_J$ to $2r_J$, but it will become clear from \S\ref{Sec:int-time} that the lower value would exclude a significant number of non-escapers, while the higher value includes such a large proportion of escapers that the selection of initial conditions} would become {very inefficient. Therefore we have settled on a compromise of $r_{ic} = 1$.}

{ It is clear that initial conditions chosen in this way are still not an equilibrium distribution: the escaping population evacuates certain regions of phase space which were initially uniformly populated (in the sense of equation~\ref{eq:microcanonical}). But we can suppose that a $\Gamma$-hypersurface is partitioned into two regions: the region of escapers evolves to leave a distribution function which is equal to zero; but the region of non-escapers is indeed approximately time-invariant, as we check in \S\ref{sec:pe-observables}.
}
 
\begin{figure}
\begin{center}
\centerline{\includegraphics[width=0.68\textwidth]{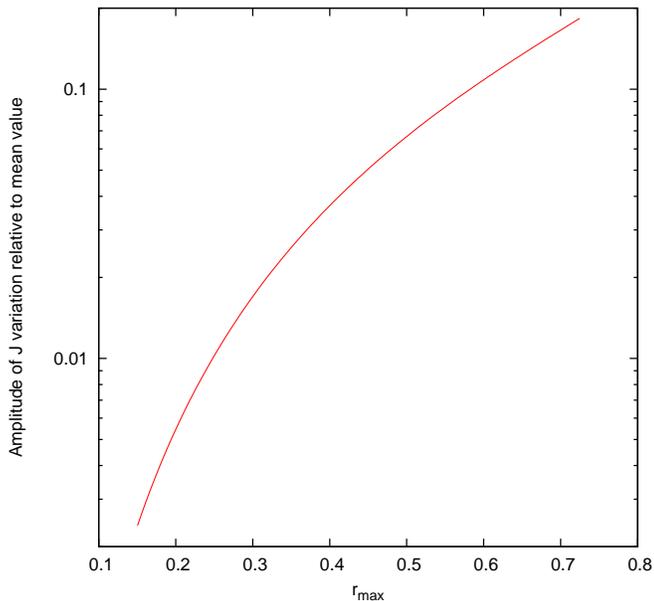}} 
\caption{Relative amplitude of the variation of angular momentum for $f$-orbits, plotted against the maximum distance from the origin. Both quantities are evaluated using the perturbation theory in Appendix~\ref{sec:appb}. In that approximation the angular momentum has a constant term and a single sinusoidal term, and the ordinate is the ratio of the amplitude of the sinusoidal term to the constant term. The orbits are planar, and so $J_N = \vert J_{zN}\vert$.} 
\label{fig:jamprel}
\end{center}
\end{figure}

\begin{figure*}
\begin{center}
\centerline{\includegraphics[width=1.1\textwidth]{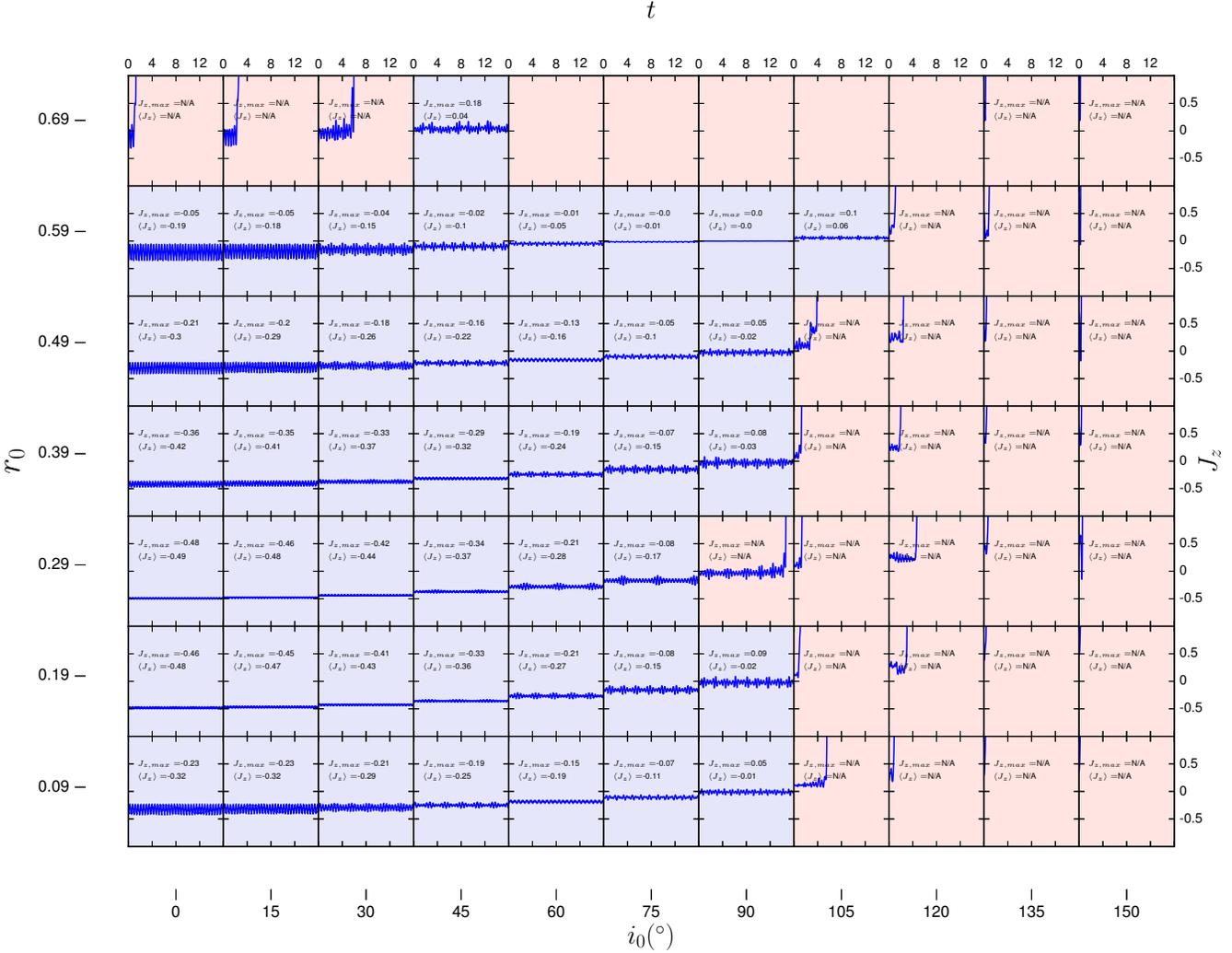}}
\caption{Time evolution of the vertical component of the angular momentum for the orbits shown in Fig.~\ref{fig:g3plus}. The orbital stability is characterised by the same colour code (red and lavender for escapers and non-escapers, respectively). For all orbits $\Gamma=3.0$ and the angular momentum is evaluated in the non-rotating frame. The initial parameters ($r_0$,$i_0$) for each orbit are given by the lower-left axis labels. The evolution of $\jnz$ as a function of time is plotted as a blue solid line, with the axes labelled upper-right (the unit of time is $2\pi$). { For non-escapers, the key in each frame gives the maximum and time-average of $\jnz$.}  {See Fig.~\ref{fig:g3plus} for a discussion of the five empty panels in the top row.}}
\label{fig:Lztableau}
\end{center}
\end{figure*}


\section{The Location of Non-Escapers in Phase Space}
\label{Sec:}

In order to construct an equilibrium distribution function that includes the phase space contribution of ``non-escapers'', as defined in \S\ref{mapeq}, 
one must identify orbits for these stars in terms of integrals of motion. In an integrable problem with three degrees of freedom we normally require three integrals. Unfortunately, the only known rigorous invariant is $\Gamma$, and so we must seek two further {\sl approximate} invariants. Both numerical results (from the previous section) and analytic ideas will guide us.

In \S\ref{sec:invariants} we present some simple arguments about approximate integrals, based on an approximate analytic theory of perturbed Kepler motion, and these ideas shape our discussion of some numerical evidence in \S\ref{sec:invariants-numeric}. Having determined suitable approximate integrals, {we next perform a number of numerical explorations, as follows. Generally, we will calculate orbits with given values of $\Gamma$ by sampling 
  the initial conditions {in the manner} 
  described in \S\ref{sec:invariantf} and Appendix \ref{sec:appa}.  In Sec.\ref{sec:basic} we begin to 
  construct a ``training set'' defined by six reference values of $\Gamma$, 
{ which coarsely}
  sample 
  the range of interest of the energy invariant $\Gamma$.}
We use these data 
to find a criterion (for {each of the basic six values of} 
$\Gamma$) which approximately discriminates escaping from non-escaping orbits, and then extend it {(by linear interpolation)} to achieve the same aim for an arbitrary value of $\Gamma$ within a wide range. { Then in Sec.\ref{sec:refinement}} {the viability of the resulting criterion is 
    tested on an independent set of orbits, characterised by a finer{, regular} grid {of} values of $\Gamma$ within the range of interest (a 44-sample ``validation set'').}  {This test reveals that some enlargement of the training set is required, and satisfactory results are found with a training set containing data for 19 discrete values of $\Gamma$.}
Further testing {on an even larger, independent library of orbits} is incorporated into \S\ref{sec:pe-observables}, where, however, the focus is on the properties of the {predicted} non-escaping population only.


\subsection{Approximate integrals of motion}
\label{sec:integrals}

\subsubsection{Analytic considerations}
\label{sec:analyticalvariation}
\label{sec:invariants}

The numerical evidence which we are about to present becomes much more intelligible in the light of some analytic ideas and calculations which we describe here.  We begin with the $f$-orbits which were discussed in \S\ref{sec:fandg}.  There it was pointed out that, for large $\Gamma$, these are Keplerian orbits, mildly perturbed by the time-dependent tidal field.   As shown in Appendix \ref{sec:appb}, their location is well described by a first-order perturbation calculation, even down to $\Gamma\simeq0.5$ or less.  This work is also instructive about the issue of approximate integrals. Thus Fig.~\ref{fig:jamprel}, which is based on the same perturbation theory (equation~\ref{eq:jz1}), shows the variation in $\jzn$ (i.e. the normal component of angular momentum in the non-rotating frame) for the family of $f$-orbits, plotted against a measure of the size of the orbit. Here, the orbits are planar, and so $J_N = \vert J_{zN}\vert$. From this we conclude that $f$-orbits which extend to the tidal radius show a variation whose relative amplitude is less than approximately 16\%. Though this might seem an unreasonably high amplitude of variation for an approximate integral of motion, it will be seen in due course (Table~\ref{tab:optima} below) that the numbers of non-escapers in the regime of small $\Gamma$ are very small.

This result encourages us to consider the possibility that invariants of perturbed Kepler motion in the non-rotating frame could serve as the approximate integrals which we seek. In this frame the energy is
\begin{equation}
E_N \equiv \frac{1}{2}\bvn^2 -\frac{1}{r} + \Phi_t.
\label{eq:en1}
\end{equation}
In our approximation, in which $\Phi_t$ is the {tidal} potential of Hill's problem (equation~\ref{eq:phit}), this expression, regarded as a Hamiltonian, is exactly the same as in quadrupole Lidov-Kozai theory (see \citealt[][\kch{\S}4.8.2]{2013degn.book.....M}, for an introduction).
In this theory, after we have averaged over the fast Keplerian motion and the slower motion of the perturber, several invariants of the averaged problem emerge. Two of these are the average Kepler energy (in the non-rotating frame), i.e. the average of $H_K$, defined in equation~(\ref{eq:hk}), and the average of the $z$-component of the angular momentum {$J_{zN}$ (see equation~\ref{eq:jzn}).}

In Lidov-Kozai theory there is a third important invariant, which is the average of the perturbation potential, i.e. $\Phi_t$ in our
notation. However, it is easy to see from equation~(\ref{eq:gamma}) that $\langle\Phi_t\rangle$ adds nothing to the set $\{\Gamma,\langle H_K\rangle,\langle J_{zN}\rangle\}$.

Lidov-Kozai theory is usually presented as a perturbation theory, the small parameter being the ratio of the semi-major axes of the ``inner'' and ``outer'' binaries.  Here we may measure the strength of the perturbation by the ratio of the perturbing acceleration $\nabla\Phi_t$ to the  Kepler acceleration, and this ratio is of order $a^3$. Thus to apply Lidov-Kozai theory we think of $a$ as small. Also, Lidov-Kozai motion has two basic frequencies, i.e. those of the two binaries, and in our context these are $a^{-3/2}$ (the Keplerian frequency) and 1 (the angular velocity of galactic motion). Then Lidov-Kozai oscillations have a frequency which is approximately in geometric progression, i.e. of order $a^{3/2}$.

\begin{figure}
\includegraphics[width=.45\textwidth]{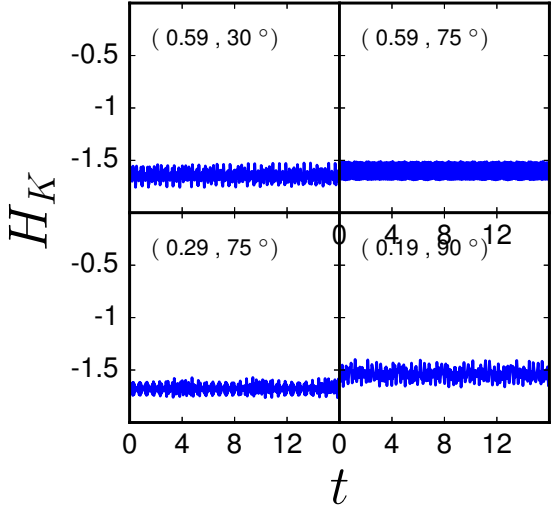}
\caption{Time evolution of the Kepler energy, as measured in the non-rotating frame, for four representative cases of non-escaping orbits selected among the numerical exploration presented in Fig.~\ref{fig:g3plus}. We highlight these cases mostly to provide evidence that the Kepler energy is an appropriate approximate constant of the motion, but also because they display a typical range of behaviour in the time variation of the magnitude of the total angular momentum (as illustrated in Fig.~\ref{fig:select4}). Values of $r_0$ and $i_0$ are given in the key of each frame, and $\Gamma=3$. Time $t$ is in units of $2\pi$.}
\label{fig:select4-hk}
\end{figure}

One last point has to be made before we turn to the numerical evidence. While $\langle\jnz\rangle$ is thought of as an invariant in Lidov-Kozai theory, it differs from the usual angular momentum by the removal of short-period oscillations, i.e. those with the frequencies of the inner and outer binaries.  For this reason we also work, not with the instantaneous values of $\jnz$ and $H_K$, but with perturbed values which correct for the main short-period oscillations. These calculations are described in Appendices \ref{sec:appc} and \ref{sec:appd}. The corrections are based on perturbation calculations, and hence are not even approximately correct when the perturbation is large, i.e. when $a$ is not small.  Therefore for practical purposes we use the corrected expressions of Appendices \ref{sec:appc} and \ref{sec:appd} only when $H_K$ is smaller than some negative cutoff value, $\hcrit$. This gives approximate invariants which we refer to below as $\hopt$ and $\jzopt$.

The corrections are smaller, in order of magnitude, for the Keplerian energy $H_K$ than for $J_{zN}$, for the following reason. When the equations for the rate of change of $H_K$ are averaged over the short-period motion, no secular term remains.  But there is a secular term for the averaged rate of change of $J_{zN}$, which has also to be averaged over the long-period motion about the galaxy before no secular term remains.  Therefore it may be expected that variations in $J_{zN}$ will be larger than those in $H_K$, as the frequency of galactic motion is generally smaller than that of Kepler motion in the domain of interest.


\subsubsection{Numerical evidence}
\label{sec:invariants-numeric}

We return to the two-parameter survey described in \S\ref{mapeq}. In Fig.~\ref{fig:Lztableau} we show the time-dependence of the $z$-component of the angular momentum in the non-rotating frame corresponding to the orbits shown in Fig.~\ref{fig:g3plus} ($\Gamma=3.0$). Each panel again has the initial position specified by the parameters $(r_0,i_0)$ given by the lower left axes.

\begin{figure}
\includegraphics[width=.45\textwidth]{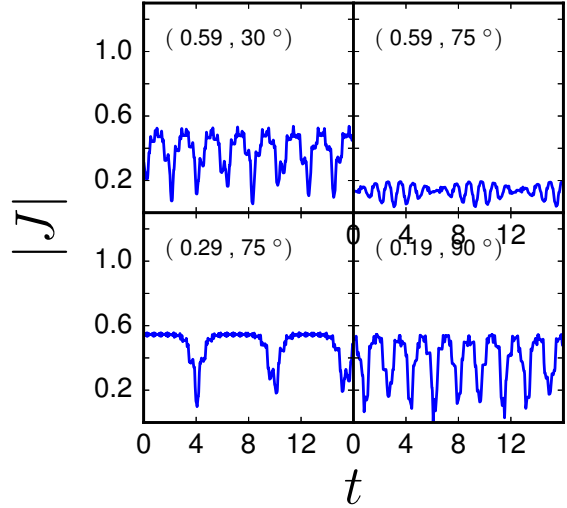}
\caption{Time evolution of the magnitude of the total angular momentum, as measured in the non-rotating frame, for the same four non-escaping orbits depicted in Fig.~\ref{fig:select4-hk}. The time evolution of the vertical component of the angular momentum of these four cases may be found in the tableau depicted in Fig.~\ref{fig:Lztableau}.}
\label{fig:select4}
\end{figure}

This tableau exhibits a number of interesting properties of the motions.  For $\Gamma=3$, the $f$-orbit would correspond to starting conditions $i_0 = 0$ and $r_0 \simeq 0.25$ \citep[][\kch{table~3}, and our \kch{Fig.~}\ref{fig:henonf}]{Henon69}, and in a wide  region of the diagram around this point the variation of $\jnz$ with time is small, i.e. it is an approximate invariant.  Towards the right-hand side of the region of non-escaping orbits, for example at $r_0 = 0.39, i_0 = 75^\circ$, $J_{zN}$ appears to exhibit an approximate periodicity, with a period which is considerably longer than the period of motion of the cluster around the galaxy (i.e. the time unit in this figure). This is a symptom of Lidov-Kozai cycling (\S\ref{sec:invariants}), which is prevalent in high-inclination orbits. The large variation of $\jnz$ for nearly-planar orbits far from the $f$-orbit, i.e. towards the top and bottom of the first two columns of the diagram, results from the short-period perturbations referred to above in \S\ref{sec:analyticalvariation}. Though this might be surprising for the orbits which start at small values of $r_0$ (which is the initial distance from the origin), Fig.\ref{fig:g3plus} shows that these are high-eccentricity orbits for which the tidal perturbation will also be large.

\begin{center}
\begin{figure*}
\begin{tabular}{ccc}
    \includegraphics[width=0.317\textwidth]{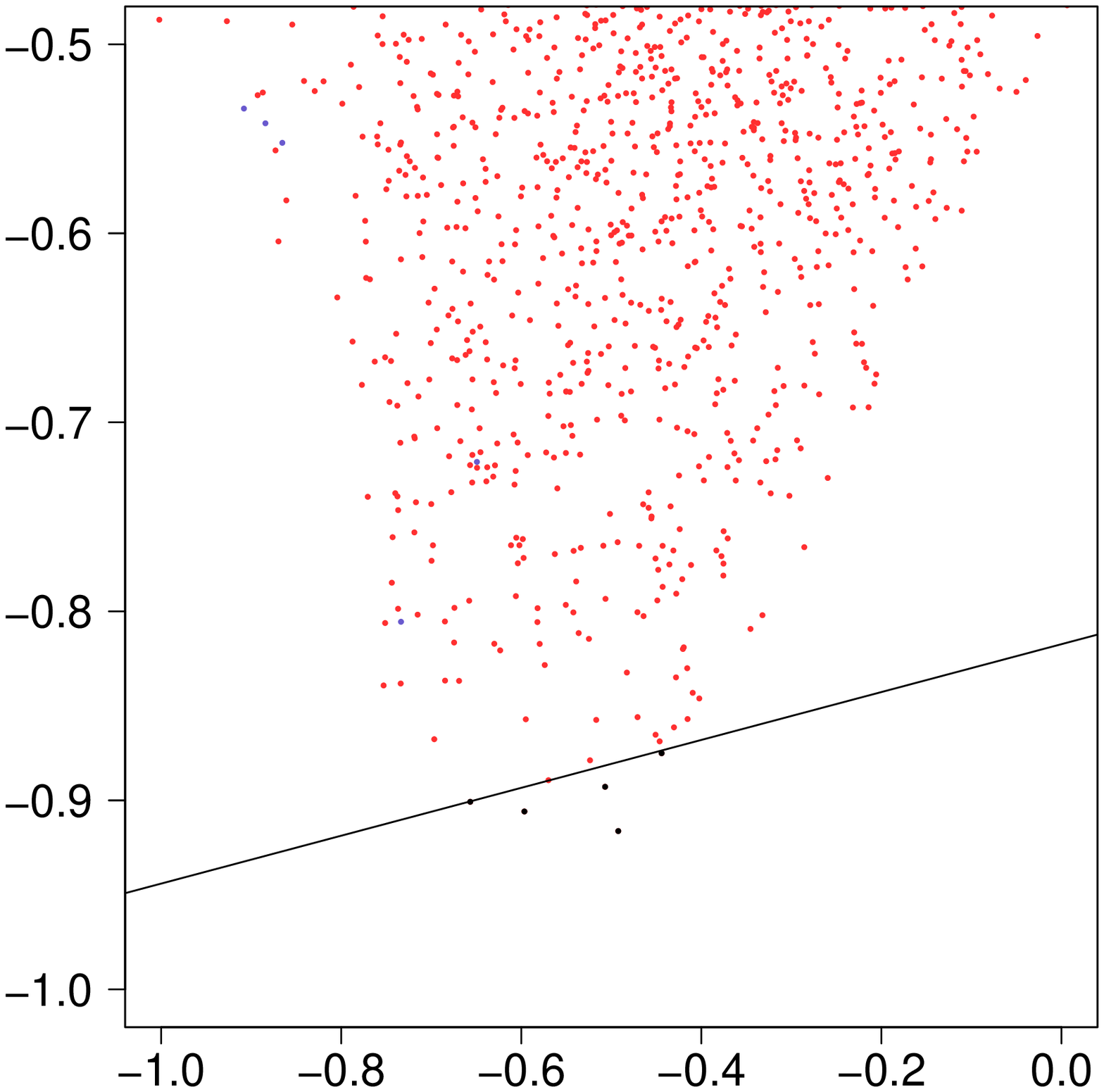} &
    \includegraphics[width=0.317\textwidth]{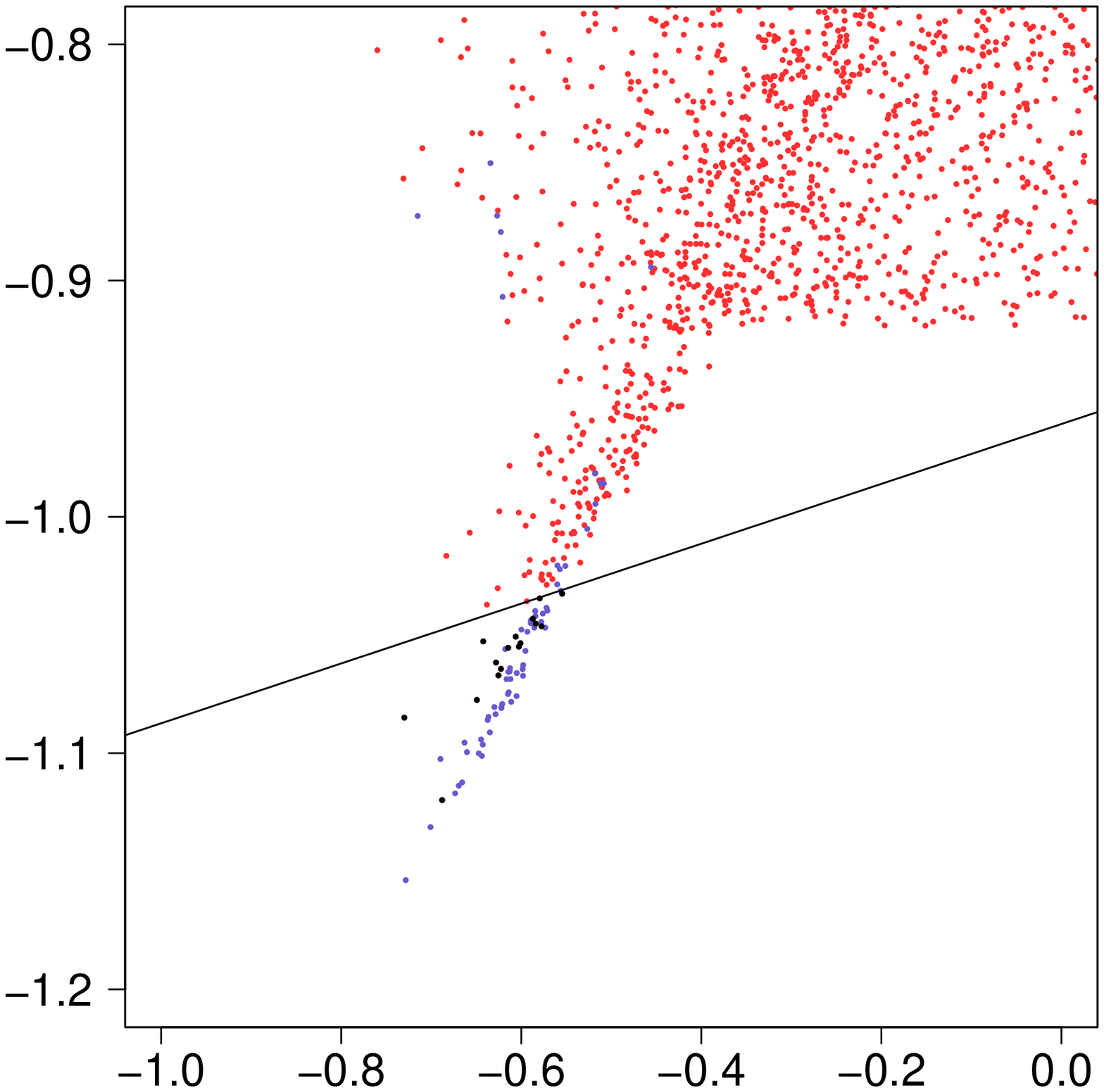} &
    \includegraphics[width=0.317\textwidth]{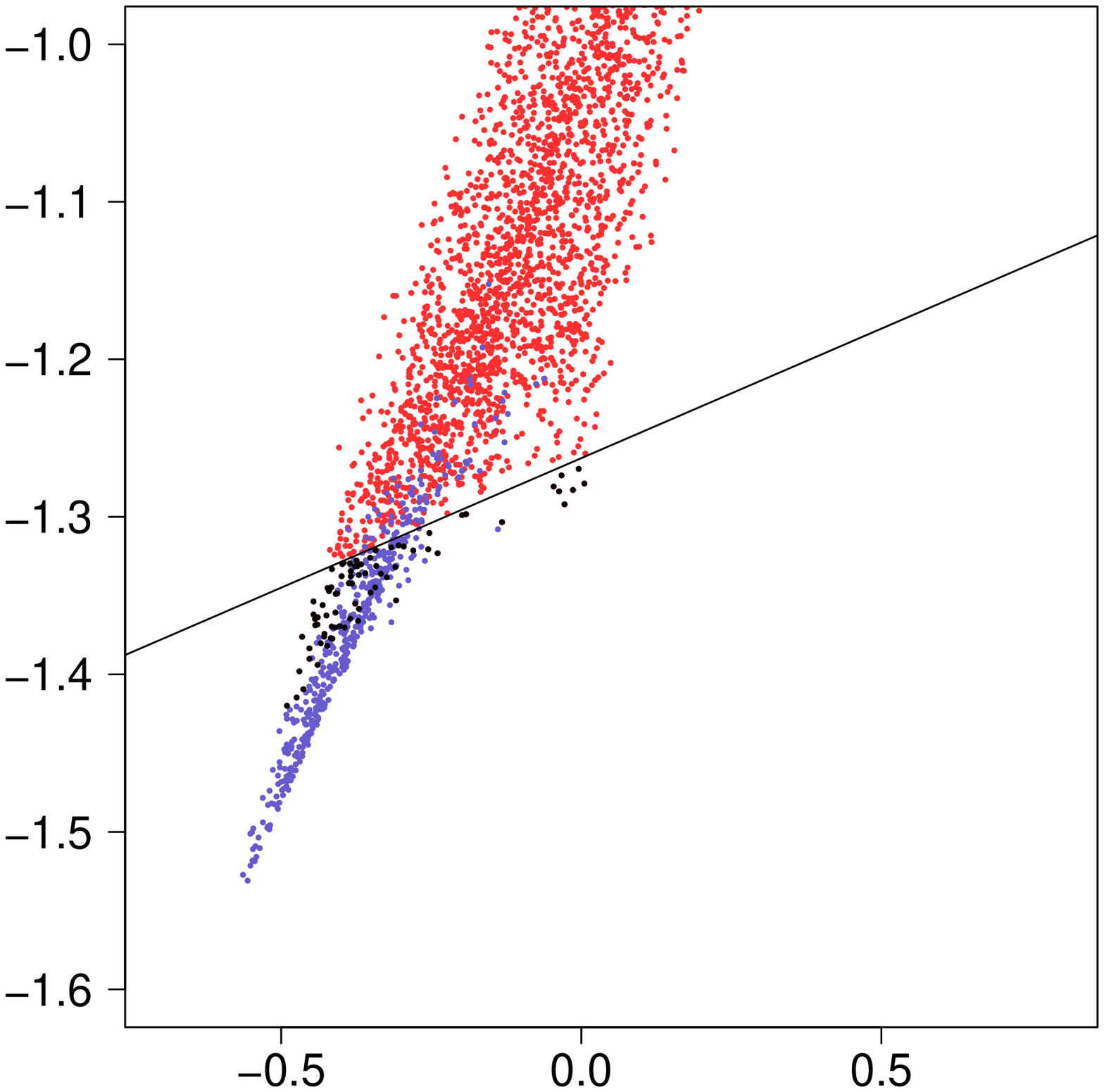} \\
    \hspace{-0.6cm} \includegraphics[scale=0.348,trim=0 0 0
-30,clip]{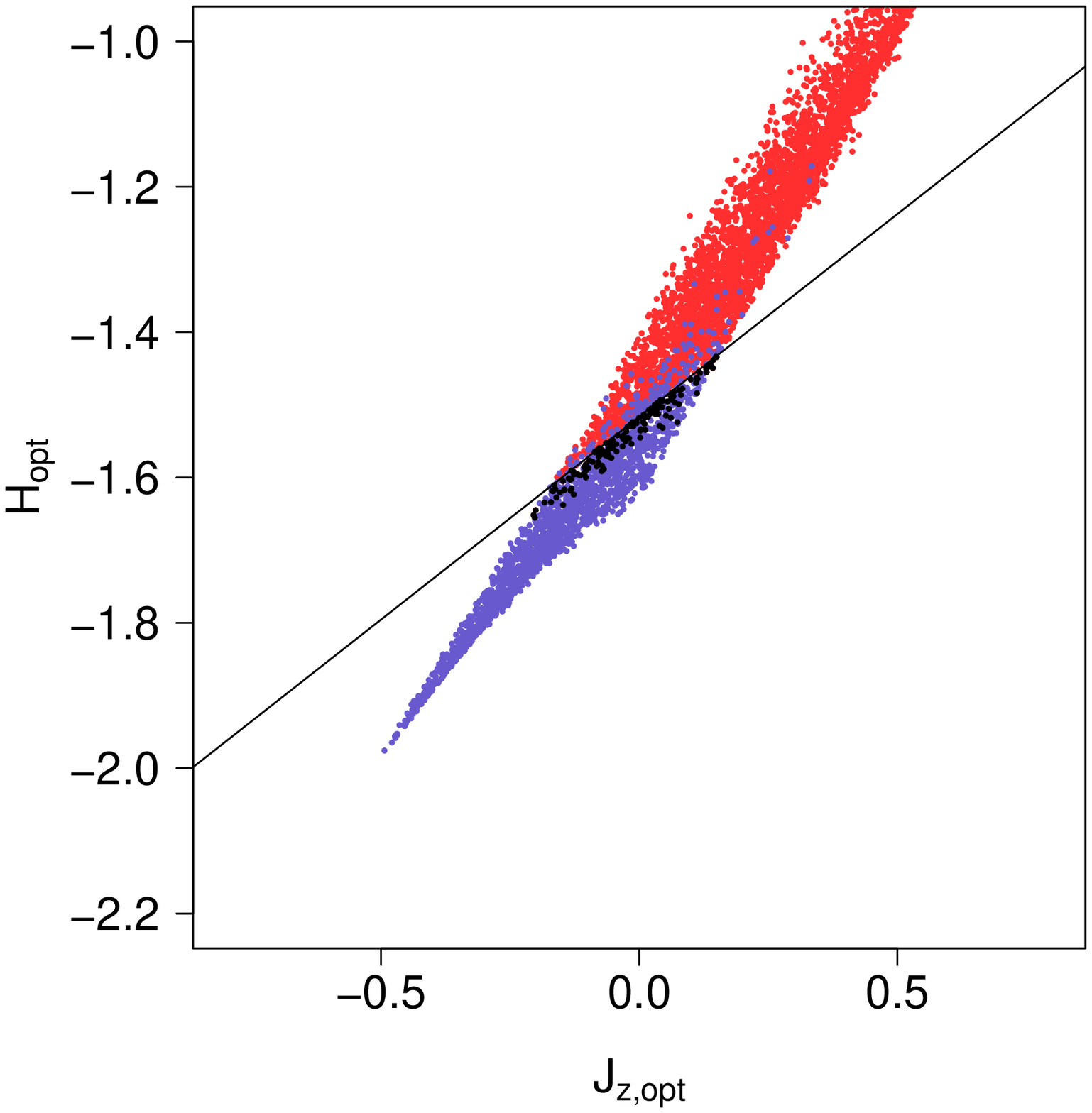}   &
     \includegraphics[trim=0 -50 0 0,clip,
width=0.317\textwidth]{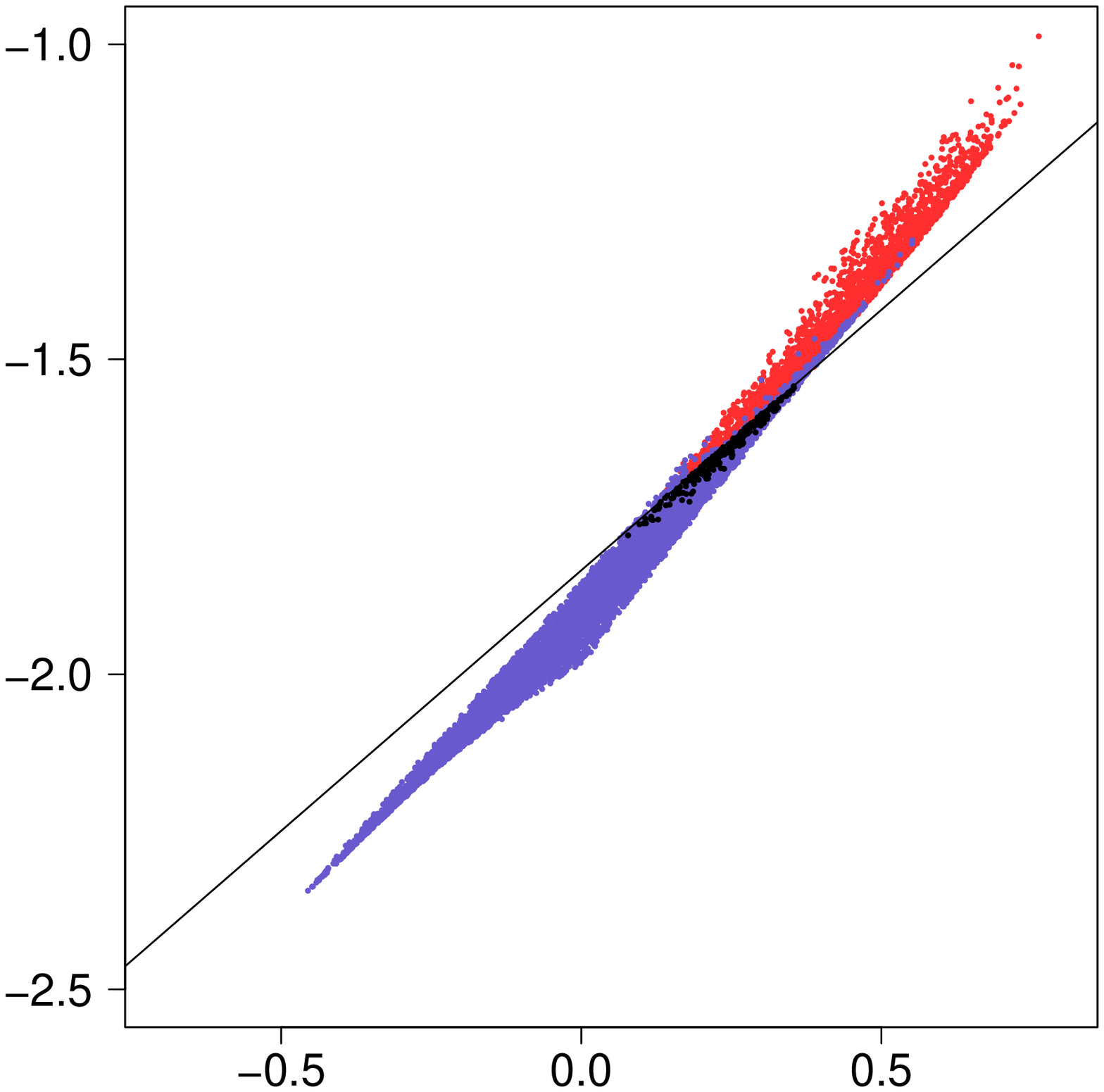} &
     \includegraphics[trim=0 -50 0 0,clip,
width=0.317\textwidth]{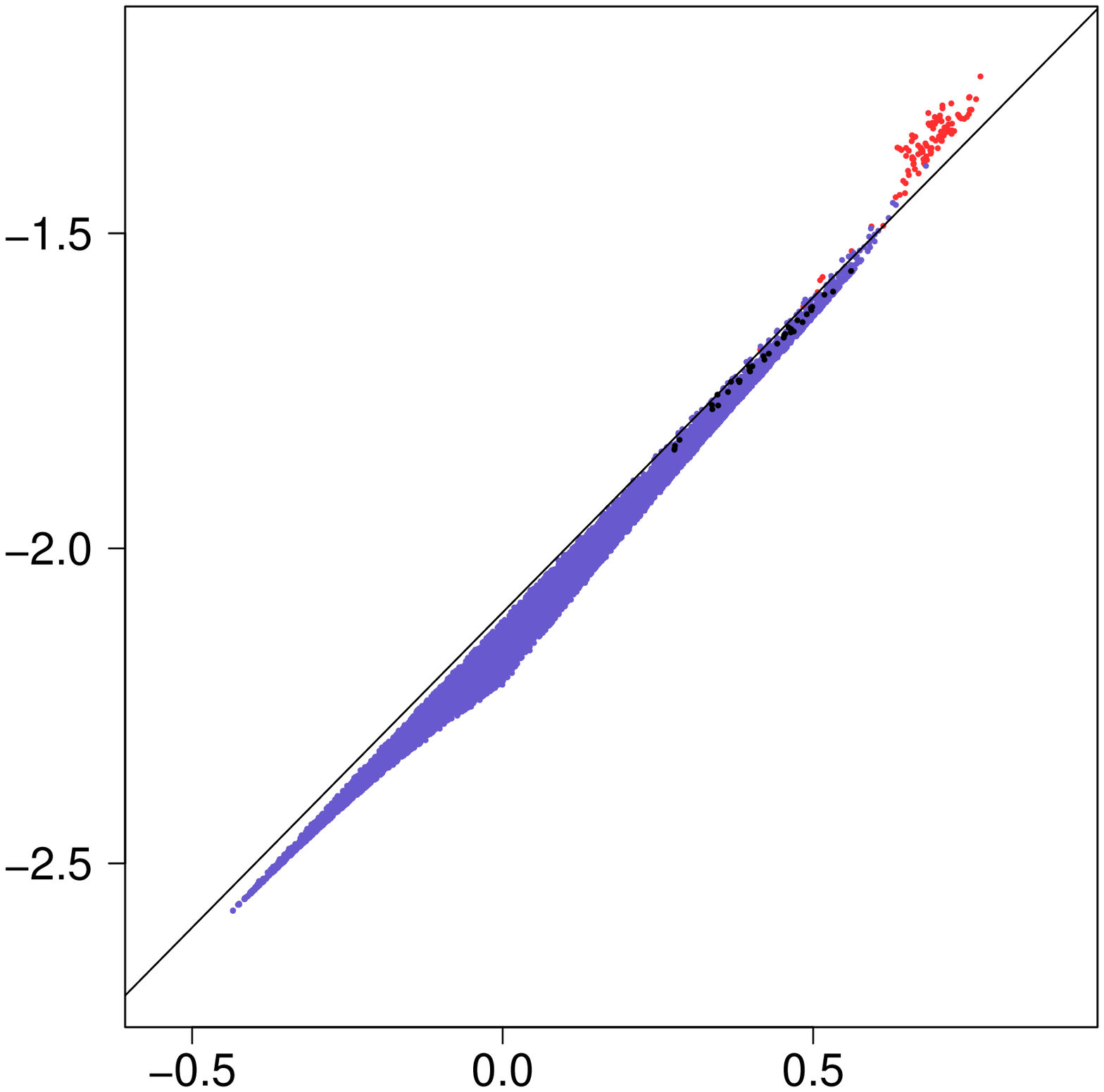}
\end{tabular}

\caption{Scatter-plots of escaping (red) and non-escaping (lavender) orbits.  The abscissa and ordinate are $J_{z, opt}$ and $H_{opt}$, respectively. The black lines give the optimal position and slope for a line dividing escapers from non-escapers (see Table~1). The black symbols give the false positives, i.e., escapers which would be incorrectly predicted to be non-escapers on the basis of this line. From top left to lower right the plots correspond to the values $\Gamma=0,1,2,3,3.8,\mathbf{4.3}$, respectively.}
\label{fig:dividing-line}\label{fig:j1-h1}
\end{figure*}
\end{center}

We have also studied the Keplerian energy in the non-rotating frame as defined in equation~(\ref{eq:hk}). For brevity we do not display all the orbits of the two-parameter survey for $\Gamma=3$, but Fig.\ref{fig:select4-hk} illustrates four examples of non-escaping orbits exhibiting typical behaviour. Though the Kepler energy exhibits oscillations, their relative amplitude is generally smaller than that of $J_{zN}$, as forecast in \S\ref{sec:invariants}, and not significantly larger in orbits which exhibit the longer-period oscillations to which attention was drawn in Fig.\ref{fig:Lztableau}. This makes the Keplerian energy an especially satisfactory quantity for inclusion in a distribution function (\S\ref{Sec:crit}).  

Finally, we have studied the time-evolution of the magnitude of the angular momentum in the non-rotating frame, $\vert J_N\vert$. Fig.\ref{fig:select4} illustrates the same four cases as in Fig.\ref{fig:select4-hk}, and exhibits clear Lidov-Kozai oscillations in all panels. The total angular momentum is not an approximate invariant of Lidov-Kozai theory, and is unsuitable for the construction of a distribution function. 


\subsection{A practical phase space criterion for non-escapers}
\label{sec:empirical}

Our purpose in the remainder of this section is to explore the distribution of motions, for a given $\Gamma$ hypersurface, which correspond to 
{ non-}escapers, and to {construct a criterion, expressed } 
in terms of approximate invariants{, which separates them from escapers}. As in \S\ref{mapeq}, our {operational definition of a non-escaper} is that, during a time of 8 revolutions of the cluster around the galaxy (i.e. a time $8\times2\pi$), the maximum distance of the star from the origin, $r_{max}$, is less than $2r_J$. Both parts of this condition require a little discussion, which we postpone to \S\ref{Sec:int-time}.

\begin{table}
\caption{Properties of the optimal dividing line between escapers and non-escapers, for {the final 19} 
  values of $\Gamma$.}
\begin{center}
\begin{tabular}{lllll}
$\Gamma$&$f_{esc}$&$N_m$&$C$&$\theta$\\
{\bf0}    &  0.9995 &   5  &  0.811  &   -0.126  \\
0.2  &  0.9994 &   4  &  1.000  &    0.214  \\
0.3  &  0.9988 &   9  &  1.064  &    0.886  \\
0.6  &  0.9975 &  14  &  1.092  &    0.679  \\
0.7  &  0.9965 &  18  &  1.011  &    0.195  \\
0.8  &  0.9953 &  20  &  0.920  &   -0.044  \\
{\bf1}    &  0.9935 &  16  &  0.953  &   -0.126  \\
1.2  &  0.9903 &  36  &  0.932  &   -0.264  \\
1.5  &  0.9832 &  57  &  1.109  &   -0.157  \\
{\bf2}    &  0.9561 &  78  &  1.246  &   -0.163  \\
2.6  &  0.8826 & 126  &  1.346  &   -0.283  \\
2.7  &  0.8672 & 128  &  1.304  &   -0.408  \\
2.8  &  0.8500 & 140  &  1.315  &   -0.440  \\
{\bf3}    &  0.8052 & 152  &  1.324  &   -0.509  \\
3.4  &  0.6683 & 287  &  1.319  &   -0.660  \\
{\bf3.8}  &  0.4615 & 251  &  1.414  &   -0.691  \\
4.1  &  0.2518 & 173  &  1.479  &   -0.723  \\
4.2  &  0.1694 & 126  &  1.552  &   -0.660  \\
{\bf4.3}  &  0.1025 &  37  &  1.486  &   -0.785  \\
\end{tabular}
\end{center}
Notes: {Boldface values of $\Gamma$ denote the core training set (Sec.\ref{sec:basic}), while the remainder were added at the stage of refinement (Sec.\ref{sec:refinement}).}  $f_{esc}$ is the fraction of escaping orbits, $N_m$ is the number of mismatches in a sample of $10,000$ (see text), and $C$, $\theta$ are the parameters of the optimal line (equation~(\ref{eq:line})) separating escapers from non-escapers in a scatter-plot of $\hopt,\jzopt$. (These are, respectively, variants of the Keplerian energy and the $z$-component of the angular momentum, both in the non-rotating frame.)  
\label{tab:optima}
\end{table}

\begin{figure}
\begin{center}
\centerline{\includegraphics[width=0.68\textwidth]{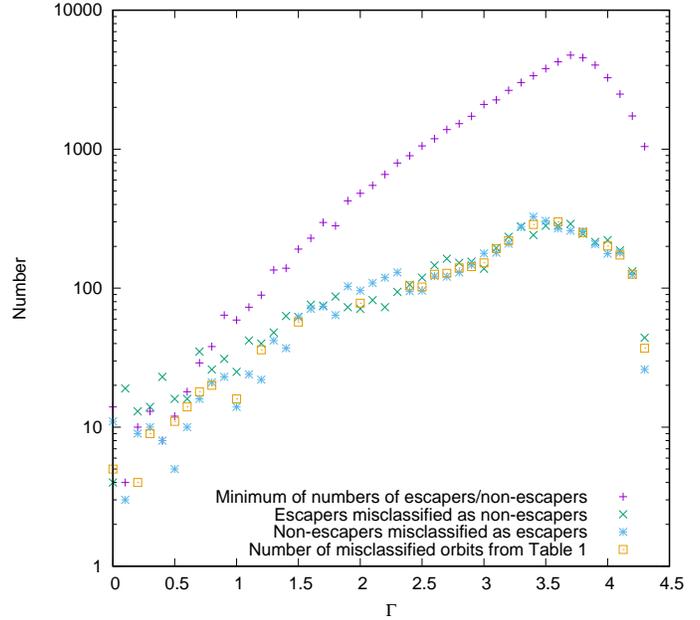}}
\caption{Mismatches of the interpolated escaper criterion when tested against {the total of} $440,000$ orbits {of the validation set}, sampled independently at 44 evenly spaced values of $\Gamma$. For comparison the minimum of the number of escapers and non-escapers (based on data in column 2 of Table 1) is plotted. Also included is data from column~3 of Table~\ref{tab:optima}. For the {validation} 
  data the numbers of mismatches (i.e. the $N_{m-}$ escapers and the $N_{m+}$ non-escapers which were classified wrongly) need not be equal, as they are in the data from Table~\ref{tab:optima}.}
\label{fig:large-survey}
\end{center}
\end{figure}

\subsubsection{Construction of a basic criterion}\label{sec:basic}

We first considered a representative set of six values of $\Gamma$, viz. $\Gamma =0,1,2,3,3.8,\alv{4.3}$, and selected initial conditions according to the recipe in Appendix \ref{sec:appa}.  For each {sample}, $10,000$ initial conditions were generated, the equations of motion (\S\ref{sec:units}) were integrated numerically for the stated time interval, and orbits divided into escapers and non-escapers according to the value of $r_{max}$. {Overall, these $6$ samples constituted the core of the ``training set'' on which we have defined the empirical discrimination criterion described below (though it was subsequently enlarged as 
  described in Sec.\ref{sec:refinement}).} The results {for the basic six values of $\Gamma$} are plotted in Fig.\ref{fig:j1-h1}, which shows scatter-plots in the plane of two putative, approximate invariants (\S\ref{sec:invariants}).

Our next task was to find an empirical way of separating escapers from non-escapers from plots such as those in Fig.~\ref{fig:j1-h1}, which illustrates our adopted solution. From visual inspection of plots such as this, we concluded that there was no obvious benefit in trying to demarcate the two kinds of orbit by anything other than a straight line, and we chose the orientation and position of this line so as to equalise the number of mismatches, $N_m$, on either side, i.e. the number of escaping orbits which lie below (or to the right of) this line, $N_{m-}$, and the number of non-escapers which lie on the other side, $N_{m+}$. The choice of optimisation criterion (i.e. $N_m$, rather than, say, the fraction of mismatches, $f_m$) requires some justification, which will be taken up in \S\ref{Sec:crit}.

For each value of $\Gamma$, the determination of the optimal dividing line {was} 
done automatically, by carrying out a search on a
relatively fine grid of values of the slope of the putative optimal line (i.e., we considered 1000 evenly spaced values of the parameter $\theta$ defined below, over the range $[-\pi,\pi]$); the position of a line with this slope {was} 
then 
advanced until the condition of equal mismatches, $N_{m-}= N_{m+}$, {was} 
first met. Note, however, that it {was} 
also necessary to choose the value of $\hcrit$ discussed in \S\ref{sec:invariants}; this was done manually, giving $\hcrit = -0.92$.  Results are also plotted in Fig.~\ref{fig:j1-h1}, and numerical values 
are {among those} given in Table~\ref{tab:optima} {, i.e. those with values of $\Gamma$ in boldface}. 
We write the optimal line as 
\begin{equation}
C + \hopt\cos\theta + \jzopt\sin\theta = 0,\label{eq:line}
\end{equation}
and values of $C$ and $\theta$ are given in the last two columns of the table. 

\subsubsection{Test and refinement of the criterion}\label{sec:refinement}

To construct a criterion valid for arbitrary $\Gamma$ in the range $[0,4^{4/3}]$, we have opted for 
linear interpolation, as attempts to construct simple fitting formulae were less successful. To test {the} viability of the resulting criterion, we have constructed an independent set of orbit data, at intervals of 0.1 in $\Gamma$ within the above range, i.e. $44$ values; for each value, $10,000$ orbits were computed. {(Overall, these 44 samples represent our ``validation set'', and they were not altered in subsequent refinement of the escape criterion.)}  Then we computed, for each {of the 44} values of $\Gamma$, the number of mismatches of both kinds, i.e. $N_{m-}$ and $N_{m+}$, obtained by comparing the actual number of escapers/non-escapers with those predicted by the linear interpolation, and again we judged the success of the criterion by considering the magnitude of the difference $\vert N_{m-} - N_{m+}\vert$.  At this point it was not possible to arrange for equality of the two numbers $N_{m-}$ and $N_{m+}$, because we were no longer free to vary the constants $C$ and $\theta$, but we used an approximate statistical criterion based on Poisson statistics to gauge whether the difference in the two numbers was acceptable. 

    {By this measure, even with linear interpolation between them, the original six values of $\Gamma$ were insufficient: it was clear that intermediate values of $\Gamma$ were poorly predicted, especially in the range where the fraction of escapers (Table~\ref{tab:optima}, col.~2) is changing rapidly with $\Gamma$.  For this 
        reason 
        we 
        extended the ``training set'' beyond the basic six values of $\Gamma$. The {resulting refinement} 
        of the ``training set'' has been performed iteratively upon revision of the resulting linear interpolation formula. This process was continued until it was judged, by the approximate statistical criterion described above\footnote{Incidentally, the statistical criterion was quite severe.  The 44-sample ``validation set'' included independent samples for all the values of $\Gamma$ which are present in Table~\ref{tab:optima}, and in one case failed to meet the criterion.}
        , that no further useful improvement could be obtained. }


The quality of the final result{, based on the 19 values of $\Gamma$ in the final training set (Table \ref{tab:optima}\footnote{  Inspection of the values in the table reveals a somewhat irregular dependence of $C$ and $\theta$ on $\Gamma$. This is certainly associated with the relatively small number of orbits ($10,000$ in each
sample), and could be improved with much larger samples.  
}),} can be seen in Fig.~\ref{fig:large-survey}, which also presents the minimum value for either the number of escapers or the number of non-escapers for the {44} values of $\Gamma$ {in the validation set}.  From study of the numbers of mismatches, it is clear that the interpolated escape criterion is not markedly worse when tested on {this} independent data, i.e. data which was not used in the creation of the interpolated criterion. The maximum number of mismatches is about 300, which is less than 10\% of the maximum number of either escapers or non-escapers.  At each end of the range of $\Gamma$ the percentage of mismatches can be much greater, but the total numbers are also much smaller; thus such mismatches can reasonably be ignored in terms of constructing a {sample of non-escapers}. 
  

\section{A model with predicted non-escapers}
\label{Sec:Exp2}

Our final objective is the construction of a model of a star cluster, and to reach this we proceed in two steps.  First we consider the {\sl observable} \alv{properties} of the {predicted non-escapers}, as defined on the basis of the {empirical discrimination criterion} elaborated in \S\ref{sec:empirical}, i.e. their surface density and kinematics. Only then do we add on an underlying bound population, to produce a synthetic approximate model of a star cluster (\S\ref{Sec:delta}).

\begin{figure}
\centerline{\includegraphics[width=0.68\textwidth]{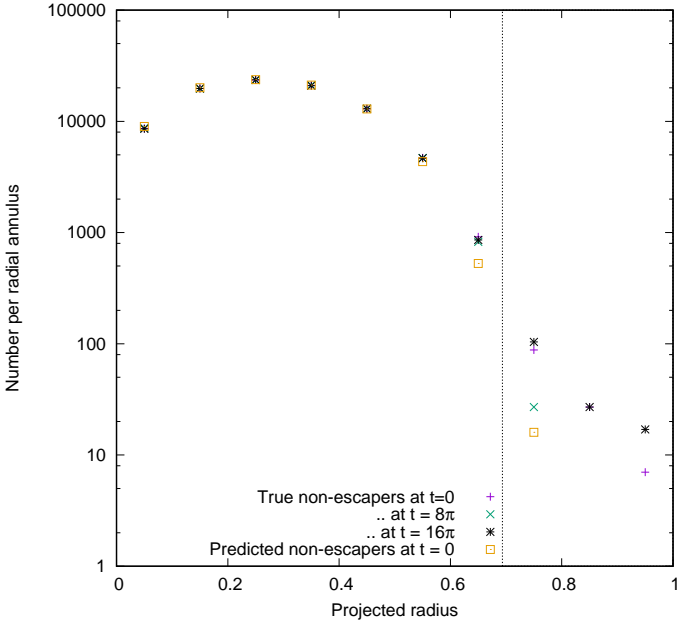}}
\caption{Projected distribution in radius of non-escapers, from a sample of 500,000 orbits. Three symbols give, for each radius, the numbers of non-escapers in the given annulus at three times: $0$, $8\pi$ and $16\pi$. The fourth symbol gives the projected numbers of initial conditions which are classified as non-escapers according to the interpolated criterion of \S\ref{sec:empirical}. The bin-width is $0.1$, and the line of sight is the $y$ direction. The vertical dotted line marks the Jacobi radius.}
\label{fig:rho}
\end{figure}


\subsection{Observable distributions of the predicted non-escapers}
\label{sec:pe-observables}

In this section we study a third sample of numerical solutions of Hill's equations, this time $500,000$ orbits with values of $\Gamma$ sampled randomly and uniformly from the range $[0,4^{4/3}]$.  For given $\Gamma$, the phase space coordinates were selected using the procedure described in \S\ref{sec:invariantf} {and Appendix \ref{sec:appa}}, including the restriction to initial positions such that $r < r_{ic}=1$. Such a sample represents an independent data-set on which the criterion elaborated in \S\ref{sec:empirical} can again be tested quantitatively. After numerical integration the orbits were classified as escapers/non-escapers, again depending on whether or not $r$ exceeds $2r_J$ at any time up to $16\pi$. By studying the distribution of the initial conditions of the non-escapers, we obtained a sample of a phase-space distribution which is confined to the domain of non-escapers, and uniform in this domain on each $\Gamma$-hypersurface.  

In this sample there were $91,597$ non-escapers, and $91,782$ non-escapers were predicted by the {empirical discrimination} criterion (\S\ref{sec:empirical}). Their spatial distribution is summarised in Fig.~\ref{fig:rho}. The first remark to be made about this figure concerns the {non-escapers}, which are plotted at three times. For most radii these three symbols closely coincide, showing that the spatial distribution is almost stationary; it was for this purpose that the microcanonical distribution of \S\ref{sec:invariantf} was adopted. The second remark is that there are some differences between the distribution of the non-escapers and those orbits predicted as non-escapers by the criterion. However, inside the Jacobi radius, this difference is not larger than the 10\% difference which is to be expected from the results of \S\ref{sec:empirical}, with the sole exception of the point just inside $r_J$. Outside the tidal radius, even though the numbers are relatively small, it is not so clear that the numbers of non-escapers are consistent with being stationary in time; indeed in the outermost two bins there are no non-escapers at $t = 8\pi$.  A possible reason for the non-stationarity is that the sample of non-escapers, defined as in \S\ref{sec:nomenclature}, includes some orbits that would escape in a longer interval of time, and that such orbits tend to have large radius. Incidentally, the scaling of the {ordinate} 
in the figure is not physically meaningful, but numbers of actual orbits are given in order that the sampling error of the points can be estimated. An alternative scaling is presented in the \S\ref{Sec:delta}.

\begin{figure}
\centerline{\includegraphics[width=0.68\textwidth]{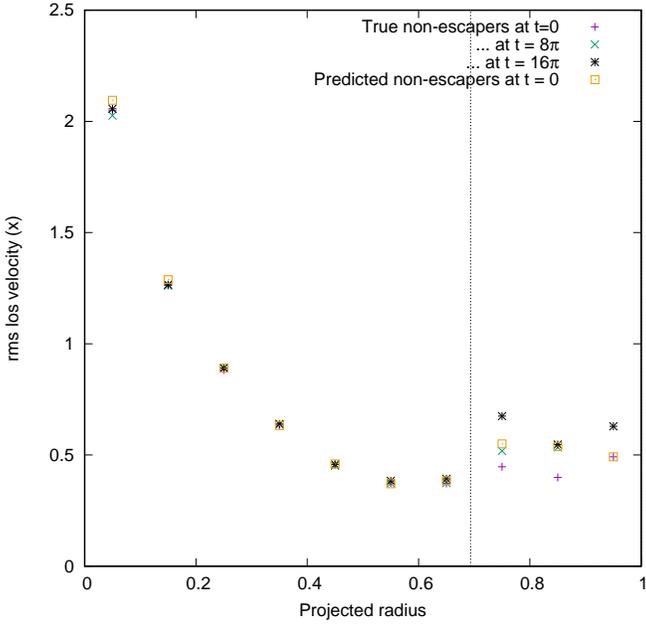}}
\caption{Distribution in projected radius of the root mean square line-of-sight velocity along the $x$-direction. From a sample of $500,000$ orbits, non-escapers are plotted at times $0, 8\pi$ and $16\pi$ (first three symbols in the key). The fourth symbol gives the result for initial conditions classified by our criterion as predicted non-escapers. The bin width is 0.1, and the vertical dotted line marks the Jacobi radius.}
\label{fig:v2x}
\end{figure}

\begin{figure} 
\centerline{\includegraphics[width=0.68\textwidth]{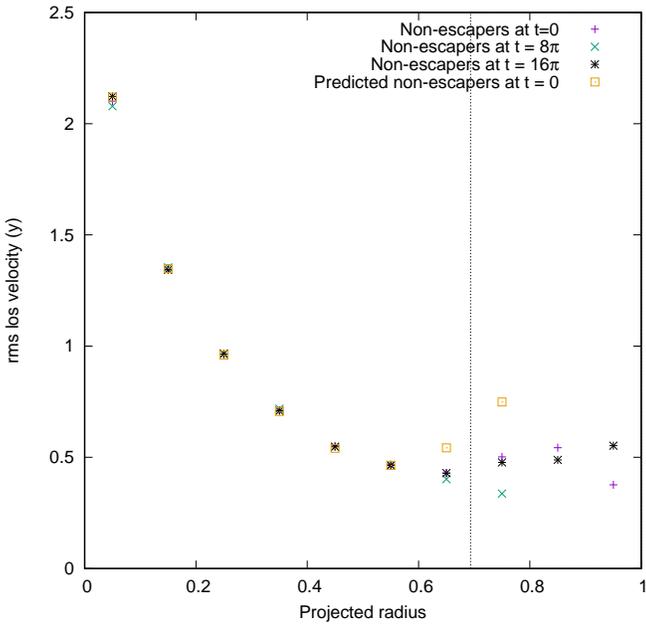}}
\caption{Distribution in projected radius of the root mean square velocity along the $y$-direction.  Other details as in the caption to Fig.\ref{fig:v2x}.}
\label{fig:v2y}
\end{figure}

The distinctive feature of potential escapers is their speed, which gives them an energy greater than the escape energy. Therefore we consider next their kinematic behaviour, in the non-rotating frame, beginning with the line-of-sight velocity dispersion. In fact we show two results, depending on whether the line of sight is in the direction of $x$ (Fig.~\ref{fig:v2x}) or $y$ (Fig.~\ref{fig:v2y}). The first of these is the line of sight towards the galactic centre at $t=0$ and each interval of $2\pi$ thereafter. Again there is little evidence of evolution with time of the non-escapers, except possibly in the last three bins. Even there, the apparent evolution with time may not be significant, as the numbers of points are very small, indeed comparable to those along the $y$ line of sight (Fig.~\ref{fig:rho}). In the interior of the cluster (in projected radius) the comparison with the results {produced by} 
the criterion {of Sec.\ref{sec:empirical}} suggests a mismatch of order 10\% at most. In the second diagram (Fig.~\ref{fig:v2y}), however, just inside $r_J$, i.e. the bin at $r = 0.65$, the mismatch is larger. Fig.~\ref{fig:v2y} gives data for the line of sight along the direction of motion of the cluster about the galaxy at the time of each evaluation shown. This includes the region close to the Lagrange points in the outermost bins, where the motions of would-be escapers are particularly complicated.  

The projected anisotropy, $\beta = 1 -\sigma_t^2/\sigma_r^2$, where the final term is the ratio of the transverse to radial components of the velocity dispersion, is shown in Figure 13. It has a distinctive form, ranging from mild transverse anisotropy {($\beta < 0$)} near the centre to strong
radial anisotropy around the tidal radius.  This is quite the opposite of what might be expected, as we have defined non-escapers by the
property that their maximum radius is bounded; therefore, near the largest radii which they reach, one might expect that they should exhibit mostly transverse motions. Nevertheless, study of individual orbits shows that non-escapers which reach radii close to $r_J$ tend to have high eccentricity, at least in projection{, and this diminishes $\sigma_t$}. The results for the three largest radii are heavily affected by the very small numbers of stars which contribute. Even so, at smaller radii the agreement between {non-escapers} and {predicted non-escapers} is little better than qualitative.

\begin{figure}
\centerline{\includegraphics[width=0.68\textwidth]{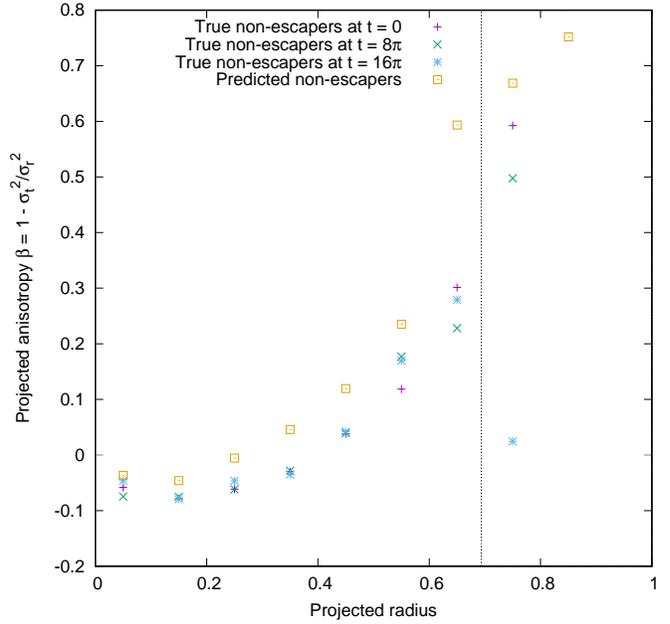}}
\caption{Profile of the projected anisotropy parameter $\beta$ (defined in the $y$-axis label in terms of the dispersion in the transverse and radial directions in the $y,z$ plane). The line of sight is the $x$-axis, and the vertical dotted line marks the Jacobi radius.}
\label{fig:aniso}  
\end{figure}

Finally we turn to an odd moment of the velocity distribution, with the mean rotational speed about the $z$-axis (Fig.~\ref{fig:vphi}). Except at large $r$ the agreement is remarkably good. Again the outer three bins are based on small numbers of orbits, but still the discrepancies between the results for {non-escapers} and predicted non-escapers around $r_J$ are too large to be explained by sampling error. 


\subsection{A complete model with predicted non-escapers}
\label{Sec:delta}

\subsubsection{Description of the model}\label{sec:model-description}

In this subsection we describe the final
goal of this paper: a
model of a star cluster in a tidal field with a population of
potential escapers.  While the model is complete in this sense, it is
not self-consistent.  Rather, we add the population of 
{ predicted non-escapers, much } 
as described in \S\ref{sec:pe-observables}, to a
self-consistent model of the bound population, but we do not include
the contribution which {the added population} 
make to the potential.
Furthermore, we assume, as in all previous parts of this paper, that
the {non-escapers} 
are moving in a Keplerian potential, which we
take as an adequate approximation to the potential of the bound
population, at least at radii where the {non-escaper} 
population
(and tidal effects generally) become significant.

Next, \kco{we assume that the value of the distribution function of the bound population, at energies just below the escape energy, equals that of the non-escaper population just above the corresponding energy.} 
{The reason for this assumption is that the population of potential escapers must be created from the bound population by any one of several processes which take them across the critical energy for escape.  One such  process is two-body relaxation, which we think of as diffusive.  Another is the slow decrease of the depth of the potential well, caused by the escape of stars from the cluster, which can be thought of as either a slow heating mechanism, or a mechanism which simply lowers the critical escape energy.  Both processes imply that the potential escapers are created by drift or diffusion across the critical energy.}

\begin{figure}
\centerline{\includegraphics[width=0.68\textwidth]{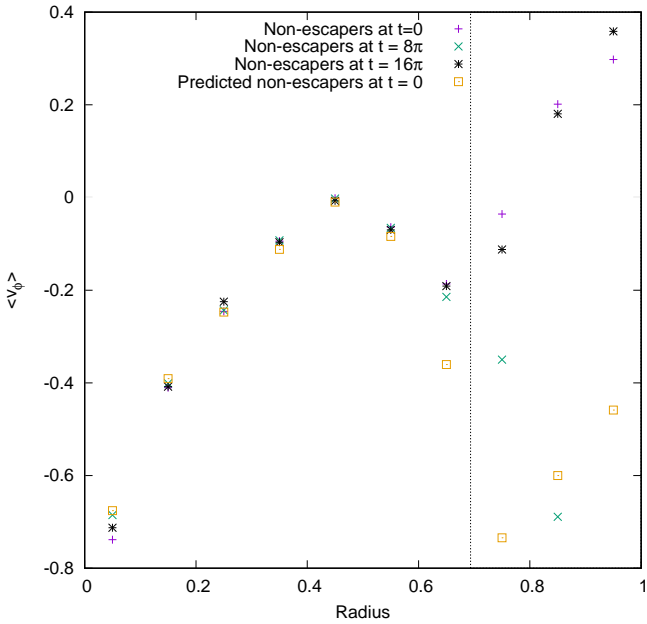}}
  \caption{Distribution in {\sl three-dimensional} radius of the  mean rotational  velocity about the $z$-axis.  Other details \dchg{are} as in the caption to Fig.\ref{fig:v2x}.}
  \label{fig:vphi}
\end{figure}

{ These arguments}
rule out
almost all of the familiar models for the bound population, such as
King and Wilson models, {in which \kco{the distribution function} $f$ tends to zero as the critical energy is approached from below.}  Indeed the only models that come to mind are
the Woolley models \citep{1954MNRAS.114..191W,WD61} and the $n = 3/2$ polytrope
\citep[see, for example,][]{2008gady.book.....B}, which has constant
$f$.  We opt for the Woolley model.

Note that our approach, {of combining a Woolley model with our results on non-escapers,} does further violence to the dynamics in
the following sense: {Woolley} 
models are built on the assumption that the
tidal field inside the cluster is zero, whereas it plays a vital role
in our theory of 
non-escapers.  Therefore the definition of
``energy'' in {the two populations} 
is different.\footnote{\kco{Incidentally, because of this difference in the definition for ``energy", the final, combined distribution function in the model is not continuous at the escape energy.}}  A further elaboration of our
approach would be to use, for the bound population, a self-consistent
model including the tidal field
\citep{HR1995,BertinVarri2008}, but in the interest
of simplicity 
we choose the Woolley model, and assume it is
Roche-lobe filling, i.e. its edge radius equals the tidal (Jacobi)
radius. 

{The Woolley model has a phase-space mass-density of the form
\begin{equation}
\kco{f_w(\br,\bv)} = \left\{
\begin{array}{ll}
  A\exp{(-2j^2(E - E_J))}&\mbox{~~if $E<E_J$}\\  
  0&\mbox{~~otherwise}
\end{array}\right.
\label{eq:woolleyf}
\end{equation}
 where $E$ is the one-particle energy per unit mass \kco{assuming the underlying potential for the Woolley model}, $E_J$ is its value at the truncation radius ($R_t$),
and $A$~and~$j^2$
are constants.  We choose the corresponding distribution function for the potential escapers, and so sample the {\sl canonical} distribution
\begin{equation}
\kco{f_{can}}(\br,\bv) = A\exp{(j^2(\Gamma-\Gamma_J))},\label{eq:canonicalf}
\end{equation}
 since 
by equation~(\ref{eqn:Gamma3D}) $\Gamma = -2E$, except for the tidal potential.
We also impose the condition $r < r_{ic} = 1$, for consistency with \S\ref{sec:invariantf}, and the condition $0<\Gamma<\Gamma_J \equiv 3^{4/3}$ introduced in Sec.\ref{mapeq}.}

\begin{figure}
  \centerline{\includegraphics[width=0.68\textwidth]{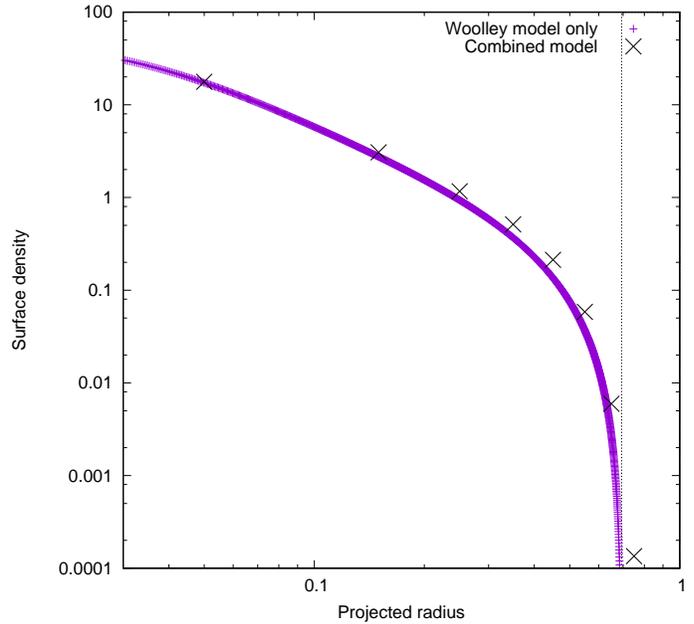}}
  \caption{Surface density profile of a $W_0 = 7$ Woolley model with
    and without potential escapers.
   The units are those of
\S\ref{sec:units}, i.e. the tidal radius is $r_J = 3^{-1/3}$ (marked by a vertical dotted line),
    the total mass (of the Woolley model) is 1, and $G = 1$.}
  \label{fig:stitched-density}
\end{figure}

\subsubsection{Practical procedure}

{
\begin{enumerate}
\item First we perform the usual numerical integrations to construct the density profile, etc, for a self-consistent Woolley model.  This calculation adopts the units of \citet{King66}.  For example, we introduce the scaled potential $W = -2j^2V(r)$, where $V(r)$ is the potential at radius $r$ whose zero-point is the truncation radius of the model, so that $E_J=0$ in eq.(\ref{eq:woolleyf}).  Then we integrate Poisson's equation in the form of 
  eq.(16) in \citet{King66} (but of course with appropriate changes in his formulae for the density, because of the different distribution function).

Next, the model must be scaled to the system of units used in  the remainder of our calculations, i.e.  the units of \S\ref{sec:units}.  There 
the cluster potential is $-1/r$, corresponding to
a cluster of mass unity {if we also set $G=1$} (and so the cluster mass is the unit of mass),
and the tidal radius is $r_J = 3^{-1/3}$.
King's units can be scaled to a model of given mass and radius by assigning values to the central density $\rho_0$ and the core radius $r_c$. 
\citet[][equation~(40)]{King66} writes the total mass 
as
\begin{equation}
  M = \rho_0r_c^3\mu,
\end{equation}
where $\mu$ is a constant depending on the scaled central potential
$W_0$.  In his units King also denotes the truncation radius 
as $R_t$, and the core radius is the unit of length.  To
carry out the required scaling to the units of \S\ref{sec:units}, then, we choose
\begin{eqnarray}
  r_c &=& 3^{-1/3}/R_t\\
\rho_0 &=& r_c^{-3}/\mu.
\end{eqnarray}

\item The second step in construction of the model is to sample a sufficient number of stars from the Woolley model and from the canonical distribution eq.(\ref{eq:canonicalf}).  The procedure for the latter is described in Appendix \ref{sec:appe}, but in the present context we sample the two parts of the model simultaneously, as follows.

\begin{figure}
  \centerline{\includegraphics[width=0.68\textwidth]{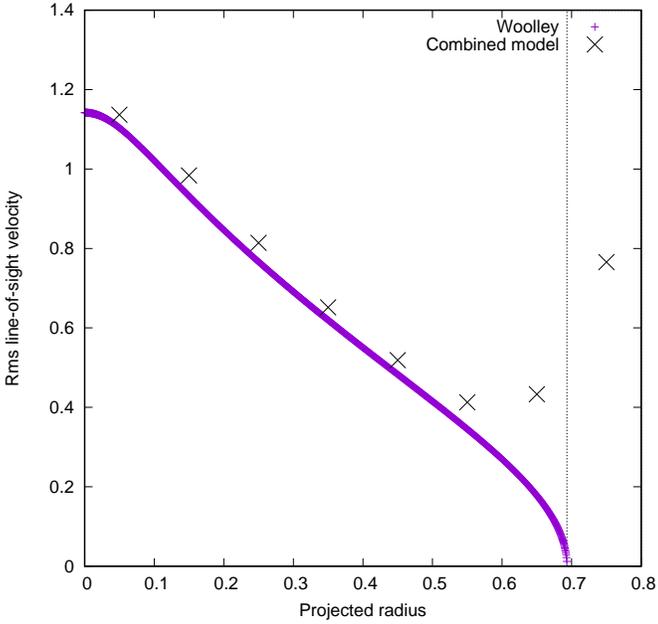}}
  \caption{Velocity dispersion profile of a $W_0 = 7$ Woolley model with
    and without potential escapers.  The units are given in the caption to Fig.\ref{fig:stitched-density}.
  }
  \label{fig:combined-vdp}
\end{figure}

  To construct a single particle, the spherical polar coordinates $(r,\theta,\phi)$ are generated with uniform density on the intervals $(0,r_{ic}), (0,\pi), (0,2\pi)$, respectively, where $r_{ic} = 1$ was introduced in Sec.\ref{sec:invariantf}.  \kco{The joint density of $r,\theta,\phi$} is given in eq.(\ref{eq:fintegral}) \kco{for the canonical distribution}, while for the Woolley model the corresponding expression is
  \begin{eqnarray}
    f_w(r,\theta,\phi) = \frac{2\pi A}{j^3}\exp{[j^2(-2\phi_w + 2\phi_w(r_t))]}\times\nonumber\\
\times    \left(\int_0^{s_{max}}s^{1/2}e^{-s}ds\right) r^2\sin\theta,
  \end{eqnarray}
  where $\phi_w$ is the potential of the Woolley model, and $s_{max} = \sqrt{-2j^2(\phi_w-\phi_w(r_t))}$.  The argument of the first exponential is simply the scaled potential $W$ in King's notation.

  Now let $F$ be a bound for $f_w + f_{can}$.  (A bound for the second contribution is easily obtained from eq.(\ref{eq:fcanmax}), while it is obvious that
  $f_w\le 2\pi Aj^{-3}\exp(W_0)\Gamma(3/2)$, where $\Gamma$ is the gamma function.)  Selecting a value of $f$ uniformly distributed in the interval $(0,F)$, we proceed as follows:
  \begin{enumerate}
  \item if $f<f_w$, the particle is a member of the Woolley population, and its velocity is easily generated from a normal distribution;
  \item if $f_w\le f\le f_w + f_{can}$, it is a member of the canonical population, and its velocity is easily generated as described at the end of Appendix \ref{sec:appe}
    \item otherwise the point $(r,\theta,\phi)$ is discarded, and a new point is selected
  \end{enumerate}
\item At the end of the generation of the sample, the particles in the canonical distribution are tested against the criterion in Sec.\ref{sec:empirical}.  In other words the values of $\hopt$ and $\jzopt$ (introduced in Sec.\ref{sec:invariants}) are calculated.  (To recap: the initial values of $H_K$ and $J_z$ are calculated; if $H_K > H_{crit} = -0.92$   (see Sec.\ref{sec:empirical}) these are adopted as $\hopt$ and $\jzopt$, respectively;  otherwise first-order perturbative corrections are calculated, as described in Appendices \ref{sec:appc} and \ref{sec:appd}, respectively, and then the corrected values are used as $\hopt$ and $\jzopt$.)  The value of the Jacobi integral is calculated, and values of $C$ and $\theta$ (not the spherical polar coordinate) are interpolated from Table \ref{tab:optima}.  Points in the canonical sample such that $C + \hopt\cos\theta + \jzopt\sin\theta > 0$ are deleted, leaving the predicted non-escapers.
  \item The sample of $N_w$ Woolley particles and $N_{pne}$ predicted non-escapers may then be used to construct projected density profiles, etc.  Since the Woolley models must give unit mass, the mass in the population of predicted non-escapers is $N_{pne}/N_w$.  As an alternative (which we adopt in the following example) one may now discard the Woolley sample, and create the contribution of the Woolley model from the first stage of the procedure, i.e. the numerically integrated Woolley model, much as one would do if there were no population of predicted non-escapers.\footnote{While this may seem to make the creation of the Woolley population unnecessary, it is used to normalise the mass of the population of predicted non-escapers.  Alternatively, this could be done by integrating eq.(\ref{eq:canonicalf}) over the sphere $r<r_{ic}$, but this is a three-dimensional integral, which might well be most conveniently estimated by a Monte Carlo integration.  In effect, the construction of the Woolley sample serves a comparable purpose.}
\end{enumerate}}

The above four steps (i)--(iv) complete the construction of the model.


\subsubsection{An example}\label{sec:example}

{Here we present results of the above procedure for one case: a Woolley model with scaled central potential $W_0 = 7$.  The complete sample prepared in step (ii) consisted of 100,000 particles, of which $N_w = 47185$ were Woolley particles and the remainder made up the canonical sample.  After deletion of the predicted escapers in step (iii), 6667 predicted non-escapers remained.  Thus the mass of this population is approximately 0.1413, in units such that the mass of the Woolley model is unity; i.e. the predicted non-escapers make up a fraction of about 0.124 of the whole.
This
value 
lies \kco{below} 
the range of values of the relative mass in {\sl potential} escapers found  in $N$-body simulations with $N = 16384$ particles in a point-mass Galactic potential
\citep[][his
figs.~11 and 12, and \citealt{Claydon+}]{2001MNRAS.325.1323B}.  
\kco{This smaller fraction is, in fact, expected given that N-body studies include all sorts of potential escapers, including several types of transients, whereas this work focuses on non-escapers only.}  

{ As described briefly in step (iv) above,}
it is straightforward to combine the surface density (say) of the predicted non-escapers
with that
of the underlying Woolley model.  An example is given in
Fig.~\ref{fig:stitched-density}.  Though the enhancement in density is
modest (though
comparable at some radii with the
relative enhancement in mass),
the effect on the velocity dispersion profile
(Fig.~\ref{fig:combined-vdp}) is more noticeable, especially close to the
tidal radius.  The last bin lies entirely outside the Jacobi radius; the value here is entirely due to the {predicted non-}
escapers, and would not be altered if their total mass were to be reduced.  Well inside $r_J$
the effect of the {predicted non-}
escapers
is a noticeable
increase in the velocity dispersion, as those stars are
 more energetic than the members of the Woolley model.  This is
 illustrated in Fig.~\ref{fig:histogram}, which displays the
 distribution of the line-of-sight velocities in the two components at
 a line of sight close to half the tidal radius.  The total rms
 line-of-sight velocity is $\sigma = \dchg{{0.652}}$, and the well-populated tails of the
 populations extend to about $2.4\sigma$ for the Woolley model, but
there they are already dominated by the {predicted non-}
escapers, which extend to at least $\dchg{2.8}\sigma$.

\section{Discussion and Conclusions}\label{Sec:DC}
  

\subsection{Discussion}\label{Sec:disc}\label{Sec:Discussion}

\subsubsection{The definition of escape and the choice of the integration time}
\label{Sec:int-time}

In \S\ref{sec:empirical} we have integrated orbits for a time of $16\pi$ (which corresponds to approximately $1.8$~Gyr at the Sun's distance from the Galactic centre), and have defined non-escape by the condition that the maximum distance from the origin, $\rmax$, is less than $2r_J$.  We consider the second point first.

\begin{figure}
  \centerline{\includegraphics[width=0.68\textwidth]{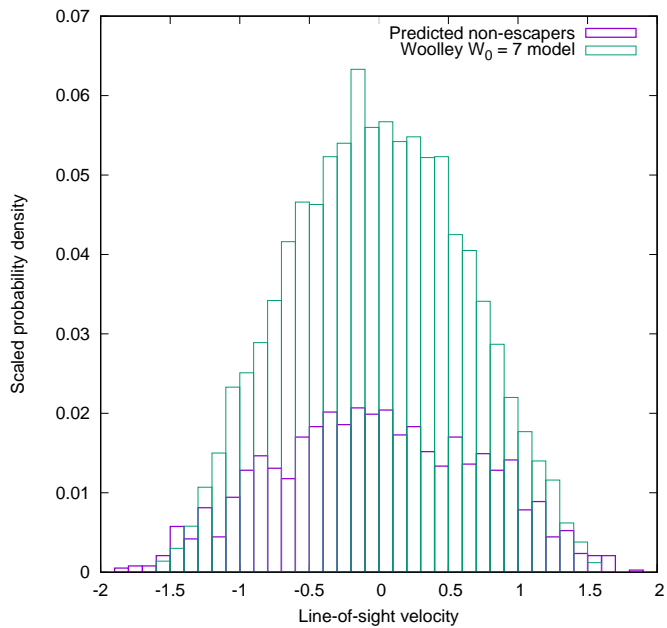}}
  \caption{Distribution of line-of-sight velocities in a $W_0 = 7$
    Woolley model and in the population of {predicted non-}
escapers.  The
    projected distance from the centre is 0.35, but for the {predicted non-}
escapers includes all stars in our sample with a projected
    distance between 0.3 and 0.4 (as in the previous two figures).
    The sample for the Woolley model is normalised to unity, and that
    for the {predicted non-}
escapers is normalised
    in proportion to their surface
    density (Fig.~\ref{fig:stitched-density}).}
\label{fig:histogram}  
\end{figure}

We illustrate in Fig.~\ref{fig:rmax} the distribution of the values of $r_{max}$, for the cases considered as non-escapers (i.e. such that $r_{max} < 2 r_J$) within the library of $500\,000$ orbits with values of $\Gamma$ sampled in the range $[0,4^{4/3}]$ (\S\ref{sec:pe-observables}).  We have also verified that, by adopting the same integration time, such a distribution of values does not depend significantly on the choice of the radial range of the initial conditions considered in our library of orbits (i.e., the value chosen for $r_{ic}$; see \S\ref{sec:invariantf}). In any case, it is clear that the limit $2r_J$ is generous, i.e. the value could be reduced considerably with very little effect on the selection of escapers.

Now we discuss the time of integration. Clearly, some non-escapers might well have escaped had we integrated for longer, though in the sample of Fig.~\ref{fig:Lztableau} it can be seen that no escapes took place between $t = 16\pi$ and $t=32\pi$. Some more general information relevant to this issue can be found in the work of \citet{2000MNRAS.318..753F}, who integrated orbits in equations like eqs.(\ref{eqn:EOMx})-
(\ref{eqn:EOMz}), but with King potentials in place of our Keplerian cluster potential, and with a different distribution on an energy hypersurface.  For a $W_0=3$ King model, their fig.~3 gives the fraction of orbits remaining as a function of time, for various values of their scaled energy $\hat E$.  Ignoring the difference between the potentials, the relation with our integral $\Gamma$ is $\hat E = 1 - \Gamma/3^{4/3}$.  They use H\'enon units, in which the Jacobi radius of their King model is approximately $3.145$. Therefore their unit of time is approximately $0.104$ times our time unit, and our integration time of $16\pi$ converts to approximately 486 H\'enon units. Their figure then shows that approximately half of orbits with $\hat E = 0.16$ ($\Gamma \simeq 3.6$) will have escaped. Despite the differences in the potential and in the distribution of initial conditions, this compares well with the result which would be inferred from the values of $f_{esc}$ in Table~\ref{tab:optima}.  Below a transition value of $\Gamma\simeq 3.4$ most orbits have escaped, while above this most have not done so.

Fukushige \& Heggie go on to show that the time scale of escape is approximately proportional to $\hat E^{-2}$. It follows that the transition value of $\Gamma$ varies with the integration time, $T$, as $(3^{4/3} - \mbox{constant}\times T^{-1/2})$. Thus we can expect that the distribution of $\Gamma$ in the population of potential escapers would slowly become narrower if $T$ were increased. In an actual stellar system, potential escapers are continually produced, with a range of $\Gamma$, and consequently a range of typical escape times.  Even as $T\to\infty$, however, the results of Fukushige \& Heggie show that a significant fraction (of order 20\%) do not escape, if the potential remains fixed.  The evolution of the population of potential escapers is a complex interplay of these processes, and beyond the scope of this paper.   It is, however, at least reasonable to presume that the orbits of potential escapers are filled from high to low $\Gamma$ (low to high $E$), since these orbits are populated diffusively from the bound population as the cluster relaxes (Sec.\ref{sec:model-description}), and by the gradual filling-in of the potential well as stars escape.

One reassuring element regarding the appropriateness of our choice of the integration time has been the {\it a posteriori} realisation that we are able to observe (when applicable) {\it at least} one full cycle of Lidov-Kozai oscillations  even for orbits near {the} threshold {of} stability {against escape}. This can be inferred from Fig.~\ref{fig:Lztableau}, but it is particularly evident in the time evolution of the amplitude of the angular momentum, and can be seen in the selection of cases illustrated in Fig.~\ref{fig:select4}.   

\begin{figure}
  \centerline{\includegraphics[width=.5\textwidth]{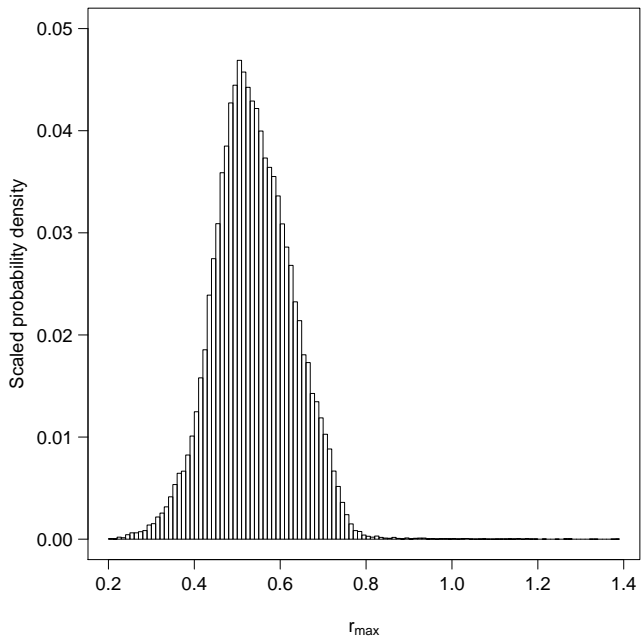}}
  \caption{Distribution of the values of the maximum distance of the star from the origin, $r_{max}$, for non-escapers (i.e. such that $r_{max} < 2\,r_J$), within the library of $500\,000$ orbits with values of $\Gamma$ sampled in the range $[0,4^{4/3}]$. The histogram is defined by 100 cells equally spaced in the range $(0,2 r_J)$, and the frequencies have been normalised to the total number of {non-escapers} 
    in the sample ($91 597$).}
  \label{fig:rmax}
\end{figure}

\subsubsection{The choice of the galactic and cluster potential}
\label{Sec:gal-pot}

We have chosen to adopt a Keplerian profile for the galactic potential in order to maintain consistency with the original numerical explorations of Hill's problem \citep{Henon69,Henon70}. In principle, other models may be considered within the same order of approximation.
Our choice of galactic potential affects the expression for the energy (Jacobi) integral through its contribution to the effective potential (see equations~(\ref{eqn:Gamma3D})~and~(\ref{eq:phit}); for discussion of the effect of different analytic models in the context of the three-dimensional Hill's problem, see, for example, Box~12.1 in \citealp{2003gmbp.book.....H} or \S2.1 in \citealp{BertinVarri2008}). 
In turn, {therefore, the choice of potential} affects the escape rate, and the size {and distribution} of the escaper and non-escaper populations
. Consequently, the ``practical criterion'' for the characterisation of escapers in phase space (described in \S\ref{sec:empirical}) is defined under the assumption of our adopted potential. \nbody\ studies \citep{Claydon+}{, on the other hand,} show that the size of the potential escaper population can differ by as much as a factor of two at comparable phases of evolution of the cluster, depending on the chosen form for the galactic potential.
 
One additional, although more theoretical, argument supporting our choice to use a point-mass potential is related to the question of the existence of the generalised concept of {Lagrange points} 
      in the context of the {\sl elliptic} Hill's problem. 
      This is guaranteed only for the Keplerian case and a small family of similar potentials (see Bar-Or et al., in preparation).            \kco{For other potentials, all that can be said is that, at peri-~and apo-galacticon, the effective potential has two saddle
points.}  

{Lastly, we note that the the field of the cluster has been approximated by that of a point mass 
  at the cluster centre.   {For most purposes} this is an inessential approximation, for 
  we argue that it is appropriate for our goal of characterising the population of potential escapers, which 
  lie outside the bulk of the cluster mass{, where the potential is nearly Keplerian}.  (In the Woolley model constructed in Sec.\ref{sec:example} the half-mass radius is approximately 0.16.)   In addition, we emphasise that, in principle, the cluster potential will not even be spherically symmetric because the tidal field is itself not symmetric (see eqs.
~\ref{eqn:EOMx}-\ref{eqn:EOMz}). Overall, partly motivated by the numerical experiments performed by \citet{Claydon+}, we consider the impact of these issues as second-order with respect to the effects determined by the choice of the galactic potential, at least outside the half-mass radius.} 

{ Predicted non-escapers do exist close to the singularity of the assumed cluster potential at $r=0$, and eq.(\ref{eq:fcanmax}) and its derivation suggest that their space density has a resulting $r^{-1/2}$ singularity there.  This would not, however, project into a cusp in the {\sl surface} density.  More serious, perhaps, is the fact that the velocity dispersion of these stars has a singularity of order $\sigma^2\sim r^{-1}$, because stars in the population of predicted non-escapers have ``energy'' $\Gamma$ in a fixed range.  Their total kinetic energy, however, is finite.  Furthermore, Figs.\ref{fig:stitched-density} and \ref{fig:combined-vdp} confirm that the singular behaviour of predicted non-escapers at small radii do not have a noticeable effect after projection and binning, but their existence should be borne in mind.}


\subsubsection{The definition of a criterion for non-escapers}
\label{Sec:crit}

As mentioned in \S\ref{sec:empirical}, there are various ways in which one might try to distinguish escapers from non-escapers entirely on the basis of initial conditions, and without numerical integration. 
In the exploratory phase of our investigation we considered several methods, which were applied to selected ``training'' data-sets (associated with six reference values of $\Gamma$ as illustrated in Fig.~\ref{fig:dividing-line}). These methods were based on various sets of criteria used for estimating either the fraction or the total number of mismatches, i.e. data that were wrongly classified according to those criteria. Since the rankings generated by different methods have been consistent with one another in all cases, the approach which we have adopted as our figure of merit is one based on the absolute number of mismatches. It was also judged that for the purpose of the final goal of this study (\S\ref{Sec:Exp2}) one may tolerate a rather large {\sl fraction} of mismatches at a given value of $\Gamma$ if the total number of mismatches is very small; this happens in the case of small $\Gamma$ (\S\ref{sec:empirical}). Thus minimising the total {\sl number} of mismatches is a better option.

As mentioned in \S\ref{sec:integrals}, several dynamical variables (such as $\jnz$~and~$H_K$) were considered in our search for a way to distinguish escapers from non-escapers. We eventually settled on the pair of variables $\jzopt$~and~$\hopt$. We did also consider for some time some variables which are based on Lidov-Kozai theory, which was already mentioned in \S\ref{sec:invariants}. These are (i) the double-averaged interaction potential $H_{LK}$ in Lidov-Kozai theory, which is actually the same as the doubly-averaged tidal potential $\Phi_t$ (equation~\ref{eq:phit}), (ii) the apocentre distance in the  Lidov-Kozai approximation, as well as in the Keplerian approximation, and (iii) the ``libration constant'' $C_{KL}$ (see \citealp{2015MNRAS.452.3610A}, equation~21), which may be used to separate the libration and rotation regimes in Lidov-Kozai theory. None appeared to offer any distinct advantage.

We also used numerical averages of several variables, such as $\jnz$, i.e. averages derived from the full numerical integration over the chosen integration time. We emphasise that such averages do not have genuine predictive power. They have been used exclusively to assess the intrinsic performance of these more ``traditional'' integrals of the motion, and to check the approximate analytic results in Appendices \ref{sec:appc} and \ref{sec:appd}.  

The first outcome of this exploratory phase of investigation was to exclude any criterion based exclusively on a single variable, as they all provided a number of mismatches which was {\it at least} $20\%$ higher than the criterion described in \S\ref{sec:empirical}. We then moved on to the exploration of several two-variable criteria, from which the selected pair $\jzopt$~and~$\hopt$ gave the best results overall, at the expense of somewhat increased complexity in the expressions to be evaluated. Finally, we have also considered a number of criteria based on three variables. The best example in this class actually outperformed our preferred two-variable criterion, but we were guided by the idea that three integrals (i.e. $\Gamma$ and two others, for which $\jzopt$ and $\hopt$ are proxy), should suffice in a problem with three degrees of freedom.

{Lastly, we wish to briefly discuss the rationale behind the selected range of values for the energy  invariant $\Gamma$. As is visible from Fig.~\ref{fig:henonfig12}, H{\'e}non's family {$f$} extends to negative values of $\Gamma$, therefore, in principle, in our study we could have considered a wider range including negative values of the energy invariant. None the less, there are two main reasons why we have decided to take into consideration exclusively positive values. First, stable orbits corresponding to negative $\Gamma$ values tend to have a significant radial extension (for reference, see the last row of Fig.~\ref{fig:f}), which would have required us to either exclude them on the basis of the operational definition of escape we have adopted, or to relax this 
  definition. A rough quantitative assessment of the first point is easily available by noting 
  the small number of non-escapers for $\Gamma \approx 0$ within our ``validation set'', as depicted in Fig.~\ref{fig:large-survey}.  Second, even just on the basis of a relatively simple tool such as the perturbation theory presented in Appendix \ref{sec:appb}, we have noticed that the validity of most of our analytic arguments, being based on perturbative approaches, tend to break down 
  for negative values of $\Gamma$ (see Fig.~\ref{fig:henonf}).}     


\subsubsection{Limitations of our model with potential escapers}
\label{Sec:lim-mod}

The dynamical model proposed in \S\ref{Sec:delta} has several limitations. In this subsection we focus on two of these.  First, although we match the bound and 
{ predicted non-}escaper population profiles using a method based on an energy continuity argument, our definition of the complete model does not include a well-posed {\sl analytic} expression for the underlying distribution function.  Though it is expressed in terms of approximate integrals of the motion, the expression is essentially numerical, i.e. interpolation in Table~\ref{tab:optima}. Second,  
our model (\S\ref{Sec:delta}) is not based on a self-consistent solution to the Poisson equation over the domain of definition. For the model of \S\ref{sec:example}, in which the proportion of 
{ predicted non-}escapers rises to approximately {14}\% of that of the bound population, this renders the results at best {approximate}
.  A self-consistent model might be achieved iteratively, by adding the density of the {predicted non-escapers} and re-solving for the potential, but then the next step would be the laborious one of adjusting the criterion (which distinguishes escapers from non-escapers) in order to take the altered potential into account.  

Despite such limitations, we consider this investigation as a first proof-of-concept, mostly in response to the practical need for a theoretical model of the potential escapers' contribution to the kinematics of the outer regions of idealised star clusters (see \citealp{Kupper10,Claydon+}), and to provide a theoretically-based understanding of their behaviour, in contrast to a purely empirical one.                   


\subsection{Conclusions}  
\label{Sec:concl}

We have reported the results of a study devoted to the construction of a dynamical model of a star cluster, in which a population of potential escapers is approximately taken into account. The starting point of this investigation is the numerical exploration of the two-dimensional Hill's problem performed by \cite{Henon69,Henon70}, and, in particular, the inspiration provided by the family $f$ of stable periodic orbits (see \S\ref{sec:fandg} and Fig.~\ref{fig:henonfig12}). 

We have extended H\'enon's picture by performing an exploration of the three-dimensional problem, through the integration of orbits starting from a much enlarged set of initial conditions (see \S\ref{mapeq} and Figs.~\ref{fig:g3plus} and \ref{fig:Lztableau}). This numerical study has guided our intuition for the development of a number of results on the approximate integrals of the motions which may be identified in this problem (see \S\ref{sec:integrals}). In this respect, a guiding principle of our search emerged from the realisation that, within a rather large range of energies, H\'enon's periodic orbits are well described as first-order perturbations of Keplerian orbits in the non-rotating frame (see Appendix \ref{sec:appb} and Fig.~\ref{fig:henonf}). In particular, this approach then suggested quadrupole Lidov-Kozai theory as an appropriate interpretative framework for the determination of suitable approximate integrals of the motion. The subsequent numerical phase of our investigation was based on a Monte Carlo exploration of the relevant energy hypersurface for selected values of the main (Jacobi) invariant (Section \ref{sec:invariantf}
).      
    
A fundamental step of our study has then been the development of a simple practical criterion for the characterisation of the potential escapers in phase space, or, loosely speaking, for determining whether an orbit is ``stable'' or not (\S\ref{sec:empirical}). {(Strictly, however, this criterion aims to identify ``non-escapers'', in a sense described in Sec.\ref{sec:nomenclature}, as a proxy for the potential escapers.)} For a given value of the Jacobi integral, we have designed this criterion as a function of two approximate invariants developed from those of Lidov-Kozai theory: the {average of the} Kepler energy and the normal component of angular momentum.  {The function} 
is 
linear 
, with coefficients tabulated {in} 
Table~\ref{tab:optima}. This condition has been determined numerically, as a result of an optimisation process to minimise the number of ``unstable'' orbits mislabelled as ``stable'' (see Fig.~\ref{fig:dividing-line}). We have then tested our criterion over a large library of orbits integrated from a random sampling of initial conditions within the entire range of energies of interest, and we have provided a characterisation of the population of {predicted non-escapers} 
in terms of several observable quantities of astronomical interest (\S\ref{sec:pe-observables} and Figs.~\ref{fig:rho}-\ref{fig:vphi}). 

Lastly, we have reached the final goal of this study: the construction of a complete dynamical model in which the previously identified population of {predicted non-}
escapers has been taken into account, together with a population of bound stars, described, as a proof of concept, by {a} Woolley {model} (\S\ref{Sec:delta} and Figs.~\ref{fig:stitched-density} and \ref{fig:combined-vdp}). We believe that such a dynamical model, although marred by a number of limitations, is the first of its kind.      	   

Our investigation was motivated by a number of intriguing results, mostly based on $N$-body simulations, which have emphasised the role played by potential escapers in shaping the structural and kinematic properties of idealised star cluster models (see especially the work by \citealp{Kupper10} and \citealp{Claydon+}). With the advent of the ``era of precision astrometry'' for Galactic astronomy, with exquisite phase space information provided by Gaia and HST proper motion studies, potential escapers may finally become {identifiable} 
in selected Galactic globular clusters, and the availability of tools to model their dynamics will be of crucial importance.          


\section{Acknowledgements}
{ We wish to thank the Referee for a particularly careful and constructive report, which especially influenced the procedure described in Sec.\ref{Sec:delta}.} We are grateful to Mario Pasquato for interesting conversations about machine learning techniques, with possible applications to our two-population discrimination problem. We are also grateful to Ian Claydon, Alice Zocchi and Mark Gieles for advanced sight of their 2016 paper, and for many conversations about potential escapers, especially during a visit to Edinburgh in July 2016. KJD acknowledges Swarthmore College, her host institution for the 2015-16 academic year, for financial support during a visit to Edinburgh that year. ALV acknowledges support from the Royal Commission for the Exhibition of 1851 and from the EU Horizon 2020 program (MSCA-IF-EF-RI 658088) in the form of research fellowships. This work was initiated during the 2014 Kavli Summer Program in Astrophysics (formerly ISIMA), hosted by CITA at the University of Toronto. We are immensely grateful to Pascale Garaud for its organisation, for financial support and, together with the other participants, for its stimulating research environment. 



\appendix


\section{Perturbation theory of $\lowercase{f}$-orbits}
\label{sec:appb}

We assume that an $f$-orbit is a  planar, tidal perturbation of a retrograde, circular Keplerian orbit in the $x,y$ plane.   We work in the non-rotating frame, and use equation~(\ref{eq:EOMN}) as the equation of motion, but ignore the $z$-component.  At lowest order we ignore the tidal perturbation, and the zero-order solution is 
\begin{equation}
\bron = a(\cos\omega t,-\sin\omega t),\label{eq:ro}
\end{equation}
where 
\begin{equation}
\omega = a^{-3/2}.\,  \label{eq:kepler3}
\end{equation}
is the orbital frequency for a Keplerian orbit. Since the tidal acceleration is of order $a^3$ smaller than the Keplerian acceleration, $a^3$ takes on the role of a perturbation parameter, though we do not write it explicitly in the first-order perturbation expansion
$\brn = \bron + \bronen$.  

Expanding equation~(\ref{eq:EOMN}) to first order, where we ignore $\bronen$ in the tidal terms, we find that the equation satisfied by $\bronen$ is
\begin{eqnarray}
{\ddot\br}_{1N} &=& - \omega^2({\bronen} - 3(\hat\br_{0N}.\bronen)\hat\br_{0N}) \nonumber \\
&&~~ +2\xor\bexr - \yor\beyr,
\end{eqnarray}
where we have used \kch{equation~(\ref{eq:kepler3})}. To solve this we resolve $\bronen$ along and orthogonal to $\bron$, writing
\begin{equation}
\bronen = \xi{\hat\br}_{0N} + \eta\hat{\dot\br}_{0N},
\end{equation}
where the two boldface vectors on the right are orthogonal unit vectors which rotate (in the non-rotating frame) with angular frequency 
$-\omega$. They form a left-handed frame. Thus the unit vectors of the original rotating frame (the ``R-frame'') are expressible as
$\bexr = \cos[(\omega+1)t]\hat\br_{0N} - \sin[(\omega+1)t]\hat{\dot\br}_{0N}$ and
$\beyr = -\sin[(\omega+1)t]\hat\br_{0N} - \cos[(\omega+1)t]\hat{\dot\br}_{0N}$.
Also, $\br_{0R}$ is expressed like equation~(\ref{eq:ro}) but with $\omega$ replaced by $\omega+1$. Routine calculations now lead to 
\begin{eqnarray}
&&  \ddot\xi - 2\omega\dot\eta-\omega^2\xi = \nonumber\\&& ~~ -\omega^2\xi +
  3\omega^2\xi + 2a\cos^2(\omega+1)t - a\sin^2(\omega+1)t\label{eq:xidd} \\
&&  \ddot\eta + 2\omega\dot\xi-\omega^2\eta =\nonumber \\&& ~~ -\omega^2\eta  + (-2a-a)\cos(\omega+1)t\sin(\omega+1)t,\label{eq:etadd} 
\end{eqnarray}
where we have avoided final simplifications to make the source of the terms clearer.

\begin{figure}
\centerline{\includegraphics[width=0.68\textwidth]{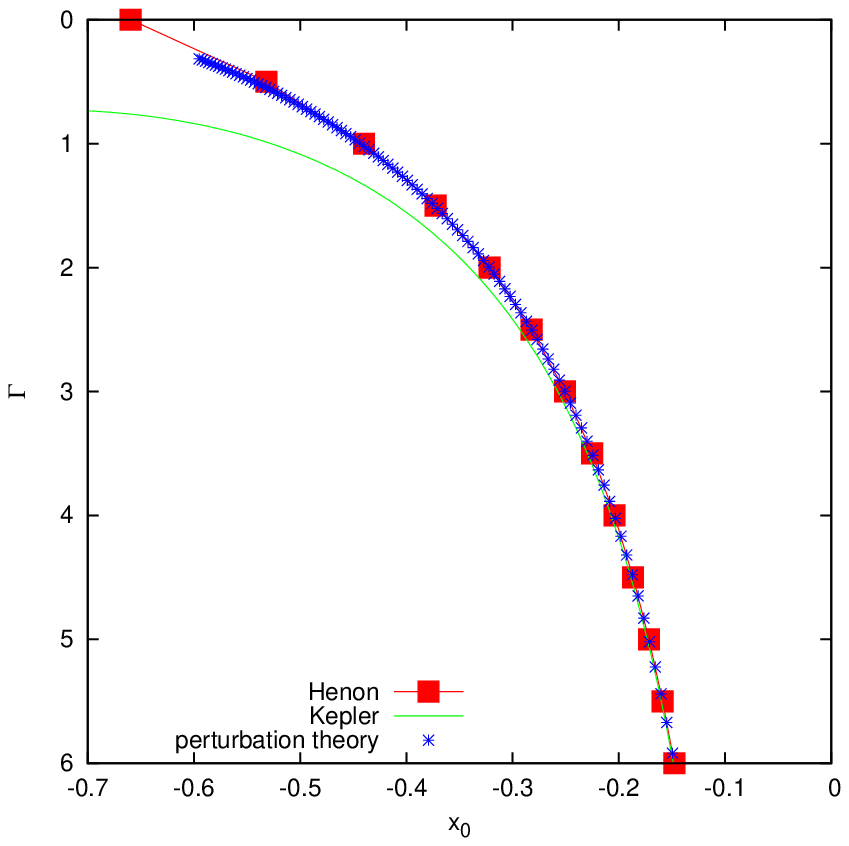}}
\caption{Family $f$ (from H\'enon 1969) compared with a Keplerian approximation and with the first-order perturbation result derived in this Appendix. For consistency with H\'enon's original study, it is assumed that the star is launched from the negative $x$-axis in the positive $y$-direction.}
\label{fig:henonf}  
\end{figure}

Simplifying and integrating \kch{equation~(\ref{eq:etadd})} gives
\begin{equation}
  \dot\eta + 2\omega\xi =
  \frac{3}{4}\frac{a}{\omega+1}\cos2(\omega+1)t + \alpha,\label{eq:etad}
\end{equation}
where $\alpha$ is a constant.  Substituting into equation~(\ref{eq:xidd}) and simplifying gives 
\begin{equation}
  \ddot\xi + \omega^2\xi =
  \frac{3}{2}\frac{2\omega+1}{\omega+1}a\cos2(\omega+1)t +
  2\omega\alpha+ \frac{1}{2}a, 
\end{equation}
with solution
\begin{equation}
  \xi =
  -\frac{3}{2}\frac{(2\omega+1)a\cos2(\omega+1)t}{(\omega+1)(\omega+2)(3\omega+2)}
  +\frac{2\alpha}{\omega} + \frac{a}{2\omega^2}.\label{eq:xi}
\end{equation}
We have ignored the kernel (complementary) function, which just moves the unperturbed motion from one circular motion to a neighbouring one.  

Substituting this result into \kch{equation~(\ref{eq:etad})} and integrating gives
\begin{eqnarray}
  \eta &=&
  \frac{3}{8}\frac{(11\omega^2+12\omega+4)a\sin2(\omega+1)t}{(\omega+1)^2(\omega+2)(3\omega+2)} -\nonumber \\
 &&~~~ - \left(3\alpha+\frac{a}{\omega}\right)t,\label{eq:eta}
\end{eqnarray}
though we have ignored a constant of integration, assuming that the perturbed solution, like the unperturbed one, starts on the $x$-axis
at $t=0$.  We choose $\alpha$ so that the secular term (proportional to $t$) vanishes. Our unperturbed motion starts at radius $a$ with
angular frequency $\omega$, but the $f$-orbit starting at this radius has a slightly different frequency, and this is responsible for the
secular term. The role of $\alpha$ is to shift the starting point to compensate. The orbit is now a periodic orbit in the rotating frame, with angular frequency $\omega + 1$.

The initial conditions of the orbit are now easily calculated by using equations (\ref{eq:etad}), (\ref{eq:xi}) and (\ref{eq:eta}), which, after transformation to the rotating frame at $t=0$, easily give
\begin{eqnarray}
  \xr &=& a - \frac{a}{6\omega^2} -
  \frac{3}{2}\frac{a(2\omega+1)}{(\omega+1)(\omega+2)(3\omega+2)}\\
{\dot x}_R &=& \yr = 0\\
{\dot y}_R &=& - \xr -\omega a + \frac{a}{6\omega} \nonumber\\
		&&~~ - \frac{3a}{4}\frac{7\omega^2+10\omega+4}{(\omega+1)(\omega+2)(3\omega+2)}.
\end{eqnarray}
Then $\Gamma$ is easily calculated. Fig.~\ref{fig:henonf} plots the results in comparison with the numerical data from H\'enon (1969) and the unperturbed (circular Keplerian) approximation.

Two further results needed in \S\ref{sec:analyticalvariation} are the angular momentum in the non-rotating frame, whose magnitude to first order is 
\begin{equation}
\vert \jn\vert = \omega a^2 + a(2\omega\xi + \dot\eta),\label{eq:jz1}
\end{equation}
which may be evaluated with the aid of \kch{equation~(\ref{eq:etad})}, and the radius (which is $r = a + \xi$ to first order), whose maximum value is easily obtained from equation~(\ref{eq:xi}).


\section{Approximate first-order perturbation theory of the Kepler energy}
\label{sec:appc} 

In the non-rotating frame the total energy is
\begin{equation}
E_N = H_K + \Phi_t,  
\end{equation}
where $H_K$ is the Kepler energy and $\Phi_t$ is the tidal potential (see {eqs.~(\ref{eq:hk}) and (\ref{eq:phit})}).  If we neglect the time-dependence of $\Phi_t$ (a point to which we return at the end of this appendix), $E_N$ is constant. Also, in first-order perturbation theory, $\Phi_t$ can be evaluated using the Keplerian approximation of the motion, which is periodic. Averaging over this period we obtain the time-averaged Keplerian energy from 
\begin{equation}
\hbar = H_K(0) + \Phi_t(0) - \langle\Phi_t\rangle.
\end{equation}
Our task in this appendix is to evaluate the last term in the right-hand side.

Since 
\begin{equation}
\Phi_t(x,y,z) = - \frac{3}{2}x^2 + \frac{1}{2}r^2,
\end{equation}
{ where we omit the subscript $R$ for the duration of this Appendix,} it is convenient to refer the Kepler orbit to the $y,z$-plane, so that $i$ (for instance) is the inclination of the plane of Kepler motion to
the $y,z$-plane.  For example we now have 
\begin{equation}
x = \sin i(\xi\sin\omega + \eta\cos\omega),
\end{equation}
where $\omega$ is the argument of pericentre, measured from the $y,z$ plane, and $\xi,\eta$ are the coordinates in the plane of Kepler motion along and orthogonal to the line of apsides, respectively.  The time averages are easily calculated as 
\begin{eqnarray}
\langle\xi^2\rangle &=& a^2\left(\frac{1}{2} + 2e^2\right)\label{eq:xi2}\\
\langle\xi\eta\rangle &=& 0\label{eq:xieta}\\
\langle\eta^2\rangle &=& \frac{1}{2}a^2(1-e^2)\label{eq:eta2}\\
\langle r^2\rangle &=& \langle\xi^2\rangle + \langle\eta^2\rangle,
\end{eqnarray}
where $a,e$ are, respectively, the semi-major axis and eccentricity. Then it follows that 
\begin{eqnarray}
\langle\Phi_t\rangle &=& \frac{1}{2}a^2\left(\frac{1}{2} +
2e^2\right)(1-3\sin^2i\sin^2\omega) +\nonumber\\
&&~~~~~~~ + \frac{1}{4}a^2(1-e^2)(1-3\sin^2i\cos^2\omega).
\end{eqnarray}

In practical terms, the calculation of $i$ and $\omega$ proceeds quite easily from the angular momentum $\bjn$ and the eccentric vector
\begin{equation}
\be = -\bjn\times\bvn - \brn/r.
\end{equation}
For example, it is easily seen that 
\begin{equation}
\cos\omega = \frac{\be.(\bexn\times\bjn)}{e\vert\bjn\vert\sin i},
\end{equation}
where $\bexn$ is the unit vector along the $x$-axis in the N frame. Note, however, that it is being assumed that the N frame instantaneously coincides with the R frame, so that the $\xn$-axis points towards the galactic centre.

The approximation that the tidal potential is static may be expected to hold for a few orbits, provided that the Keplerian frequency is
much larger than unity (the angular velocity of the tidal potential in the N frame), i.e. if $a\ll\rt$, where $\rt$ is the tidal radius.  For longer intervals of time the motion can be approximated by Lidov-Kozai theory, in which the mean Keplerian energy is constant. Thus $\hbar$ is an approximate integral of the problem in this limit. Unfortunately, we are often obliged to adopt these approximations even for values of $a$ which are comparable to $\rt$. 


\section{Approximate first-order perturbation theory of the $\MakeLowercase{z}$-component of angular momentum}
\label{sec:appd}

From \kch{equations~(\ref{eq:jzn})~and~(\ref{eq:EOMN})} we deduce that
\begin{eqnarray}
  && {\dot J}_{zN} = x_N\ddot y_N - y_N\ddot x_N =\nonumber \\ 
  && =(x_N\be_{yN} - y_N\be_{xN}).(2x_R\be_{xR} - y_R\be_{yR} - z_R\be_{zR}),
\end{eqnarray}
where $\be_{xN},\be_{yN}$ are unit vectors of the non-rotating frame.  At first we neglect the relatively slow rotation of the R-frame, and assume (as at $t=0$) that the two frames coincide.  In this approximation
\begin{equation}
{\dot J}_{zN} = -3x_Ny_N.  \label{eq:jzndot}
\end{equation}
Since the period of a Keplerian orbit is of order $a^{3/2}$, where $a$ is the semi-major axis, it follows that short-period oscillations in $\jzn$ will be of order $a^{7/2}$.

Now we average over this ``fast'' motion, following much the same route as in Appendix \ref{sec:appc}, except that we refer the Keplerian orbit to the axes of $x_N,y_N,z_N$ in the conventional way.  Thus, for example,
\begin{eqnarray}
x_N & = &\xi(\cos\omega\cos\Omega - \sin\omega\sin\Omega\cos i)+ \nonumber \\
&&~~ + \eta(-\sin\omega\cos\Omega - \cos\omega\sin\Omega\cos i),
\end{eqnarray}
where $\xi,\eta$ do have the same meaning as in Appendix \ref{sec:appc}, i.e. coordinates in the Keplerian plane.  After averaging and simplifying, we obtain the result that
\begin{equation}
  \dot{J}_{zN} = A\cos2\Omega + B\sin2\Omega,\label{eq:jnzdot}
\end{equation}
where
\begin{eqnarray}
  A &=& -3\sin\omega\cos\omega\cos i(\langle\xi^2\rangle - \langle\eta^2\rangle)\\
  B &=& -\frac{3}{2}\langle\xi^2\rangle(\cos^2\omega - \sin^2\omega\cos^2i) -\nonumber\\
     &&~~~~~ -\frac{3}{2}\langle\eta^2\rangle(\sin^2\omega - \cos^2\omega\cos^2i),
\end{eqnarray}
and the averages are exactly as in equations~(\ref{eq:xi2})~and~(\ref{eq:eta2}); we have also made use of equation~(\ref{eq:xieta}).

Now we reinstate the rotation of the R-frame.  For  a Keplerian orbit fixed in space, this is easily achieved by noting that the rotation of the axes corresponds to a decrease of $\Omega$, also with unit angular velocity.  Thus $\Omega = \Omegao - t${, where $\Omegao$ is the initial value}.  Integration of equation~(\ref{eq:jnzdot}), with initial value $\jzn(0) = J_{zN0}$, gives
\begin{eqnarray}
\jzn &=& J_{zN0} + \frac{1}{2}(A\sin2\Omegao - B\cos2\Omegao) - \nonumber \\
 &&~~~ - \frac{1}{2}(A\sin2\Omega - B\cos2\Omega).
\end{eqnarray}
Thus the average of $\jzn$, which is what we require, is given by the first half of the right-hand side {(the terms with zero subscripts)}.  The second half consists of oscillating terms with zero average.  Their amplitude is of order $a^2$, and thus much bigger (for small semi-major axis) than the high-frequency terms which we have ignored.
                                                                                                                         

\section{Sampling of the $\Gamma$-hypersurface}
\label{sec:appa}

{In the present Appendix we consider a $\Gamma$-hypersurface in phase space, but the only other restriction we apply is to the initial radius. Bearing in mind also our aim of constructing an equilibrium distribution in phase space, we note from {Jeans} Theorem that this can be done by choosing any function of $\Gamma$.}

{For these reasons we begin by considering the invariant distribution (the ``microcanonical'' distribution)
\begin{equation}
f(\brr,\bvr) = \delta\left(\Gamma + \bvr^2 - \frac{2}{r} - 3x_R^2 + z_R^2\right),
\label{eq:microcanonical}
\end{equation}
where $\delta$ denotes the Dirac delta, and we have used equation~(\ref{eqn:Gamma3D}). It follows that the marginal distribution of $\brr$ is 
\begin{equation}
f(\brr) = 2\pi\sqrt{-\Gamma + \frac{2}{r} + 3x_R^2 - z_R^2}
\label{eq:f}
\end{equation}
when the argument of the square root is non-negative.  Thus in spherical polar coordinates the space distribution is
\begin{equation}
f(r,\theta,\phi) = 2\pi\sqrt{-\Gamma + \frac{2}{r} + 3x_R^2 - z_R^2}\,r^2\sin\theta
\label{eq:f-spherical}
\end{equation}
where $x_R = r\sin\theta\cos\phi$ and $z_R = r\cos\theta$.}		

{Clearly this distribution function is non-zero for arbitrarily large $\vert x_R\vert$, and cannot be normalised.  We also expect that non-escapers (at fixed $\Gamma$) will be confined to a bounded region in configuration space, and this is 
  checked numerically in 
  \S\ref{Sec:int-time} and Fig.\ref{fig:rmax}. For these reasons we impose a further restriction on the domain of $f(\brr)$, which is the condition $r < r_{ic}$, where $r_{ic}$ is to be chosen. It is true that there are non-escaping orbits at arbitrarily large radii (Fig.~\ref{fig:henonfig12}), but these require arbitrarily negative values of $\Gamma$, which we are excluding by the restriction to $\Gamma > 0$. It follows from equation~(\ref{eq:f-spherical}) that, in the domain $r < r_{ic}$, 
\begin{equation}
f(r,\theta,\phi) \le 2\pi\sqrt{2 + 3 r_{ic}^3}\,r_{ic}^{3/2}.
\label{eq:fmax}
\end{equation}
Thus the spherical polar coordinates can be found from a simple rejection procedure, and then $\bvr$ has magnitude given by the square root expression in equation~(\ref{eq:f}) and uniformly distributed direction.}


\section{Sampling the canonical distribution}
\label{sec:appe}

Here we present a procedure for sampling the canonical phase-space density
\begin{equation}
f_{can}(\br,\bv) = A\exp{(j^2(\Gamma -\Gamma_J))}
\end{equation}
in the domain 
\begin{equation}
r < 1, 0 < \Gamma   \equiv -v^2 + \frac{2}{r} + 3x^2 - z^2 < \Gamma_J,
\end{equation}
where $A, j^2$ are constants.  {Note that the limit on $r$ means that we assume $r_{ic}=1$ in Sec.\ref{sec:invariantf}.}  As for the microcanonical distribution discussed in \kch{\S}\ref{sec:invariantf} {and Appendix \ref{sec:appa}}, we begin with the marginal distribution
\begin{equation}
  f_{can}(\br) = 4\pi \int_{v_{min}}^{v_{max}} A\exp(j^2(-v^2 - 2\Phi - \Gamma_J)) v^2 dv,\label{eq:fbr}
\end{equation}
where $\Phi =- 1/r -3x^2/2+z^2/2, v_{min} = \sqrt{\max(0,-2\Phi-\Gamma_J)}$ and $v_{max} = \sqrt{-2\Phi}$.  (Note that $\Phi< 0$ if $r<1$.)  With the substitution $s = j^2v^2$ we find that
  \begin{equation}
    f_{can}(\br) = \frac{2\pi A}{j^3}\exp\{j^2(-2\Phi - \Gamma_J)\}\int_{s_{min}}^{s_{max}}s^{1/2}e^{-s}ds,
  \end{equation}
  where the limits have their obvious meaning.  Thus in spherical polar coordinates we have
  \begin{eqnarray}
    f_{can}(r,\theta,\phi) &=& \frac{2\pi A}{j^3}\exp\{j^2(-2\Phi - \Gamma_J)\} \nonumber\\
    &&~~ \times\left(\int_{s_{min}}^{s_{max}}s^{1/2}e^{-s}ds\right)r^2\sin\theta.\label{eq:fintegral}  
  \end{eqnarray}
In general the integral is easily evaluated in terms of an incomplete gamma function. In cases where $\vert\Phi\vert$ is very large, however, (near the origin), $s_{min}$ and $s_{max}$ are also large, and care needs to be taken in the evaluation of the integral, for instance by introducing a variable of integration $t = s - s_{min}$, and developing a suitable simple asymptotic form.

To use this to select $\br$ by an acceptance-rejection procedure, we need to estimate a bound for $f_{can}$.  In fact, returning to equation~(\ref{eq:fbr}), we have
\begin{eqnarray}
f_{can}(\br) &=& 2\pi\int_0^{\min(-2\Phi,\Gamma_J)}Ae^{j^2(\Gamma - \Gamma_J)}(-\Gamma-2\Phi)^{1/2}d\Gamma\\
&\le& 2\pi\int_0^{\Gamma_J}A(-2\Phi)^{1/2}d\Gamma\\
&=& 2\pi A\Gamma_J\sqrt{\frac{2}{r} +3x^2-z^2}.\label{eq:fcanmax}
\end{eqnarray}
Then we proceed to spherical polar coordinates, as in Appendix \ref{sec:appa}.

The position vector $\br$ having been obtained, it is straightforward to sample the speed $v$ from the distribution $v^2\exp(-j^2v^2)$ in the range $(v_{min},v_{max})$.  Then the direction of $\bv$ is chosen isotropically.

\label{lastpage}


\begin{thebibliography}{99}
\bibitem[\protect\citeauthoryear{Aarseth}{2003}]{Aarseth03} Aarseth S.~J., 2003, Gravitational $N$-Body Simulations, Cambridge: Cambridge University Press
\bibitem[Antognini(2015)]{2015MNRAS.452.3610A} Antognini J.~M.~O., 2015, MNRAS, 452, 3610 
\bibitem[\protect\citeauthoryear{Barmby et al.}{2002}]{Barmby2002} Barmby P., Holland S.,  Huchra J.~P., 2002, AJ, 123, 1937 
\bibitem[\protect\citeauthoryear{Baumgardt}{2001}]{2001MNRAS.325.1323B} Baumgardt H., 2001, MNRAS, 325, 1323 
\bibitem[\protect\citeauthoryear{Bellazzini et al.}{2015}]{Bellazzini2015} {Bellazzini M., Mucciarelli A., Sollima A. et al.,\ 2015, MNRAS, 446, 3130} 
\bibitem[\protect\citeauthoryear{Bertin \& Varri}{2008}]{BertinVarri2008} Bertin G., Varri A.~L., 2008, ApJ, 689, 1005
\bibitem[\protect\citeauthoryear{Binney \& Tremaine}{2008}]{2008gady.book.....B} Binney J., Tremaine S., 2008, {Galactic Dynamics}, Princeton: Princeton University Press
\bibitem[\protect\citeauthoryear{Chandrasekhar}{1942}]{Chandra} Chandrasekhar S., 1942, Principles of stellar dynamics, Chicago: The University of Chicago Press
\bibitem[\protect\citeauthoryear{Claydon et al.}{2017}]{Claydon+} Claydon I., Gieles M., Zocchi A., \kco{2017}, MNRAS, in press
\bibitem[\protect\citeauthoryear{Correnti et al.}{2011}]{Correnti2011} {Correnti M., Bellazzini M., Dalessandro E. et al.,\ 2011, MNRAS, 417, 2411 }
\bibitem[\protect\citeauthoryear{Da Costa}{2012}]{DaCosta2012} Da Costa G.~S., 2012, ApJ, 751, 6 
\bibitem[\protect\citeauthoryear{Davoust}{1977}]{Davoust77} Davoust E., 1977, A\&A, 61, 391
  \bibitem[\protect\citeauthoryear{Drukier et al.}{1998}]{Drukier1998} Drukier G.~A., Slavin S.~D., Cohn H.~N. et al., 1998, AJ, 115, 708 
\bibitem[\protect\citeauthoryear{Drukier et al.}{2007}]{Drukier2007} Drukier G.~A., Cohn H.~N., Lugger P.~M. et al., 2007, AJ, 133, 1041 
\bibitem[\protect\citeauthoryear{Elson et al.}{1987}]{Elson1987} {Elson R.~A.~W., Fall S.~M., Freeman K.~C.,\ 1987, ApJ, 323, 54 }
\bibitem[\protect\citeauthoryear{Fukushige \& Heggie}{2000}]{2000MNRAS.318..753F} Fukushige T., Heggie D.~C., 2000, MNRAS, 318, 753 
\bibitem[\protect\citeauthoryear{Gieles \& Zocchi}{2015}]{GielesZocchi2015} Gieles M., Zocchi A., 2015, MNRAS, 454, 576 
\bibitem[\protect\citeauthoryear{Giersz et al.}{2013}]{Giersz13} {Giersz M., Heggie D.~C., Hurley J.~R., Hypki A.,\ 2013, MNRAS, 431, 2184} 
\bibitem[\protect\citeauthoryear{Gilmore et al.}{2012}]{Gilmore2012} Gilmore G., Randich S., Asplund M. et al., 2012, The Messenger, 147, 25 
\bibitem[\protect\citeauthoryear{Gomez-Leyton \& Velazquez}{2014}]{Gomez14}Gomez-Leyton Y.~J., Velazquez L., 2014, Journal of Statistical Mechanics: Theory and Experiment, 4, 6
\bibitem[\protect\citeauthoryear{Grillmair et al.}{1995}]{Grillmair1995} {Grillmair C.~J., Freeman K.~C., Irwin M., Quinn P.~J.,\ 1995, AJ, 109, 2553} 
\bibitem[\protect\citeauthoryear{Gunn \& Griffin}{1979}]{GG79}Gunn J.~E., Griffin R.~F., 1979, AJ, 84, 752
\bibitem[\protect\citeauthoryear{Harris et al.}{2002}]{Harris2002} Harris W.~E., Harris G.~L.~H., Holland S.~T.,  McLaughlin D.~E., 2002, AJ, 124, 1435
\bibitem[\protect\citeauthoryear{Heggie}{2001a}]{Heggie2001} Heggie D.~C., 2001a, in Steves B.A., Maciejewski A.J., eds, The Restless Universe. IoP Publishing, Bristol, p.~109 
\bibitem[\protect\citeauthoryear{Heggie}{2001b}]{2001ASPC..228...29H} Heggie D.~C., 2001b, ASPC, 228, 29 
\bibitem[Heggie \& Hut(2003)]{2003gmbp.book.....H} Heggie D.,  Hut P., 2003, The Gravitational Million-Body Problem: A Multidisciplinary Approach to Star Cluster Dynamics, Cambridge: Cambridge University Press 
\bibitem[\protect\citeauthoryear{Heggie \& Ramamani}{1995}]{HR1995} Heggie D.~C., Ramamani N., 1995, MNRAS, 272, 317 
\bibitem[\protect\citeauthoryear{H\'enon}{1969}]{Henon69} H\'enon M., 1969, A\&A, 1, 223
\bibitem[\protect\citeauthoryear{H\'enon}{1970}]{Henon70} H\'enon M., 1970, A\&A, 9, 24
\bibitem[\protect\citeauthoryear{Hunter}{1977}]{Hunter77} Hunter C.,\ 1977, AJ, 82, 271 
\bibitem[\protect\citeauthoryear{Johnston et al.}{1999}]{Johnston99} {Johnston K.~V., Sigurdsson S., Hernquist L.,\ 1999, MNRAS, 302, 771 }
\bibitem[\protect\citeauthoryear{Keenan \& Innanen}{1975}]{1975AJ.....80..290K} Keenan D.~W., Innanen K.~A., 1975, AJ, 80, 290 
\bibitem[\protect\citeauthoryear{King}{1966}]{King66}King I. R., 1966, AJ, 71, 64
\bibitem[\protect\citeauthoryear{K{\"u}pper et al.}{2008}]{Kupper08} K{\"u}pper A.~H.~W., MacLeod A.,  Heggie D.~C., 2008, MNRAS, 387, 1248 
\bibitem[\protect\citeauthoryear{K{\"u}pper et al.}{2010}]{Kupper10}K{\"u}pper A.~H.~W., Kroupa P., Baumgardt H., Heggie D.~C., 2010, MNRAS, 407, 2241
\bibitem[\protect\citeauthoryear{Kuzma et al.}{2016}]{Kuzma2016} {Kuzma P.~B., Da Costa G.~S., Mackey A.~D., Roderick T.~A.,\ 2016, MNRAS, 461, 3639} 
\bibitem[\protect\citeauthoryear{Leon et al.}{2000}]{Leon2000} {Leon S., Meylan G., Combes F.,\ 2000, A\&A, 359, 907} 
\bibitem[\protect\citeauthoryear{McLaughlin \& van der Marel}{2005}]{MvdM05} McLaughlin D.~E., van der Marel R.~P., 2005, ApJS, 161, 304
\bibitem[\protect\citeauthoryear{Merritt}{2013}]{2013degn.book.....M} Merritt D., 2013, Dynamics and Evolution of Galactic Nuclei, Princeton: Princeton University Press
\bibitem[\protect\citeauthoryear{Meylan, Dubath \& Mayor}{1991}]{MDM91}Meylan G., Dubath P.,  Mayor M., 1991, ApJ, 383, 587
\bibitem[\protect\citeauthoryear{Miocchi et al.}{2013}]{Miocchi2013} {Miocchi P., Lanzoni B., Ferraro F.~R. et al.,\ 2013, ApJ, 774, 151} 
\bibitem[\protect\citeauthoryear{Olszewski et al.}{2009}]{Olszewski2009} {Olszewski E.~W., Saha A., Knezek P. et al.,\ 2009, AJ, 138, 1570} 
\bibitem[\protect\citeauthoryear{Read et al.}{2006}]{Read06}{Read J.~I., Wilkinson M.~I., Evans N.~W., Gilmore G., Kleyna J.~T.,\ 2006, MNRAS, 366, 429} 
\bibitem[\protect\citeauthoryear{Renaud \& Gieles}{2015}]{RenaudGieles2015} Renaud F.,  Gieles M., 2015, MNRAS, 448, 3416 
\bibitem[\protect\citeauthoryear{Ross et al.}{1997}]{Ross97} {Ross D.~J., Mennim A., Heggie D.~C.,\ 1997, MNRAS, 284, 811} 
\bibitem[\protect\citeauthoryear{Sollima et al.}{2009}]{Sollima09} {Sollima A., Bellazzini M., Smart R.~L. et al.,\ 2009, MNRAS, 396, 2183} 
\bibitem[\protect\citeauthoryear{Sollima \& Mastrobuono Battisti}{2014}]{Sollima14} {Sollima A., Mastrobuono Battisti A.,\ 2014, MNRAS, 443, 3513} 
\bibitem[\protect\citeauthoryear{Spitzer}{1987}]{1987degc.book.....S} Spitzer L., 1987, Dynamical Evolution of Globular Clusters, Princeton: Princeton University Press
\bibitem[\protect\citeauthoryear{Takahashi \& Portegies Zwart}{2000}]{Takahashi00} {Takahashi K., Portegies Zwart S.~F.,\ 2000, ApJ, 535, 759} 
\bibitem[\protect\citeauthoryear{Wilson}{1975}]{Wilson75}Wilson C.~P., 1975, AJ, 80, 175
\bibitem[\protect\citeauthoryear{Woolley}{1954}]{1954MNRAS.114..191W} { Woolley R.~v.~d.~R., 1954, MNRAS, 114, 191}
\bibitem[\protect\citeauthoryear{Woolley \& Dickens}{1961}]{WD61} Woolley R.~v.~d.~R., Dickens R.~J., 1961, Royal Greenwich Observatory Bulletins, 42, 291
\end{thebibliography}
\end{document}